\documentclass[12pt]{article}

\usepackage{dcolumn}
\usepackage{bm}

\usepackage{units}
\usepackage{graphicx}
\usepackage{epstopdf}
\usepackage{hyperref}
\usepackage{color}
\usepackage[utf8]{inputenc}
\usepackage{cases}
\usepackage{lineno}
\usepackage{caption}
\usepackage{authblk}

\author{E.~Cazzola}
\affil{Department of Mathematics, University of Surrey, Guildford, United Kingdom}

\author{D.~Curreli}
\affil{Nuclear, Plasma and Radiological Engineering University of Illinois at Urbana-Champaign, Illinois, USA}

\author{G.~Lapenta}
\affil{Centre for mathematical Plasma Astrophysics, Department of Mathematics, KULeuven, Leuven, Belgium}
\vspace{10pt}

\begin{document}
 \title{On magnetic reconnection as promising driver for future plasma propulsion systems}

%
%
%
%
%
\date{}
\maketitle

\begin{abstract}

This work presents a more detailed analysis of the process of magnetic reconnection as promising ion beam accelerator mechanism with possible applications
in laboratory plasmas and, more importantly, in the plasma propulsion field. In a previous work, an introductory study on this subject was already carried out, 
yet under the adoption of relevant approximations, such as the limitation to 2.5D simulations and the especially use of Hydrogen plasma as a propellant, 
whose element is rarely considered in the real scenario. 
\textcolor{black}{Also, the analysis mainly focussed  on studying the physical content of the outcomes, by leaving out the analysis of
more important engineering quantities,}
such as the mass flow and thrust effectively reached out of such systems.
With this work, we intend to fill these gaps in order to provide further 
insights into the great potentiality of a future technology based on magnetic reconnection.
Additionally, one of the possibly limiting  features was the inevitable symmetric outflow produced by the reconnection process. 
Among all the possible solutions adoptable, we propose here a solution based on the particle behavior undertaken in entering the reconnection region 
according to the initial density profile. 
We demonstrate that a noticeable net thrust value can be achieved by setting up a longitudinal asymmetric density profile with a relevant drop gradient.

\end{abstract}

\section{\label{sec:introduction} Introduction}

\begin{figure}
 \centering
 \includegraphics[scale=.4]{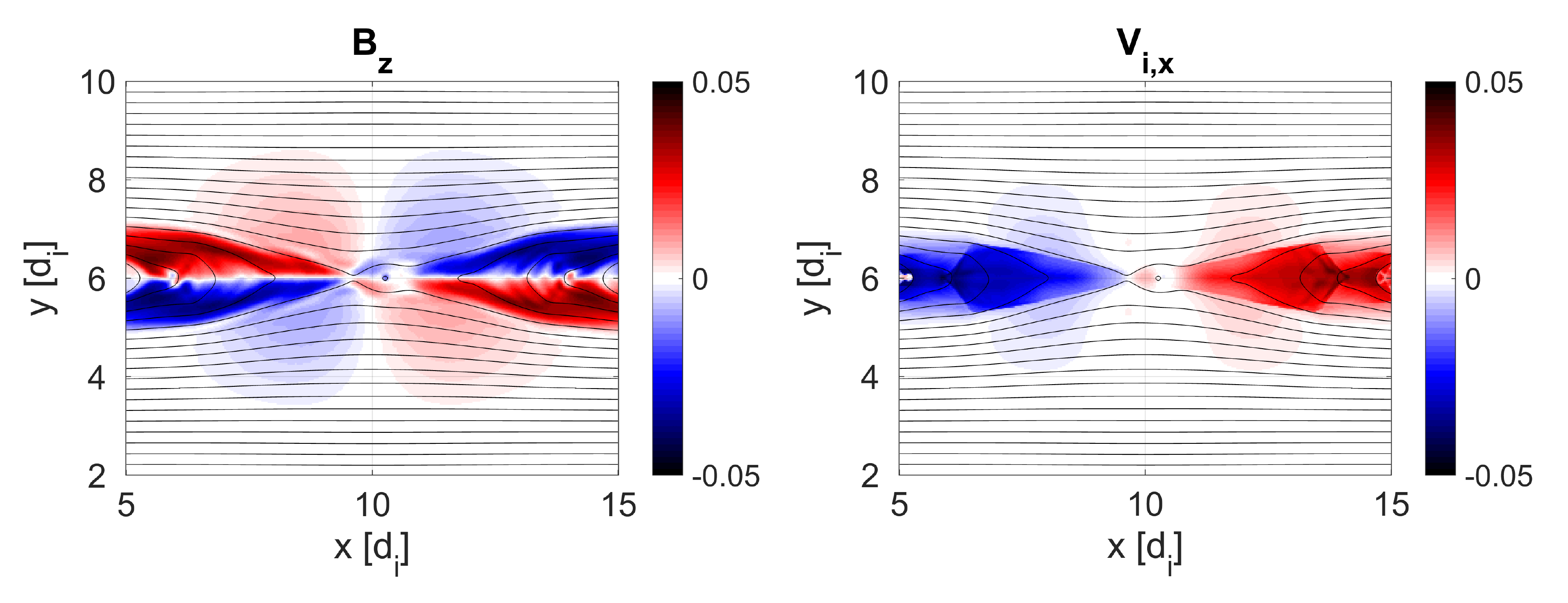}
 \caption{Representation of the main typical magnetic reconnection outcome. The left panel represents the out-of-plane magnetic field generated from 
 an initial in-plane $X-Y$ field profile. The right panel shows the $X$-component of the ion velocity at the same time step.
 \textcolor{black}{The quantities are normalized to, respectively, the initial asymptotic magnetic field $\mathbf{B}_0$ and the light-speed.}}
 \label{fig:0}
 \end{figure}

In a previous work the process of magnetic reconnection was investigated as possible efficient ion beam accelerator, by looking in particular 
at possible innovative 
 plasma propulsion applications \cite{cazzola2016b}. 
 \textcolor{black}{In short, magnetic reconnection is an electromagnetic process occurring  whenever two anti-parallel 
magnetic field lines encounter, which  causes a complete restructuration of 
the local 
 magnetic topology
 and brings to the release of a great amount of the stored 
magnetic energy (e.g. \cite{biskamp00}).
In magnetized plasmas, depending on the initial configuration, it is believed that nearly up to $50\%$ of this released
energy goes to
particles, either in form of heat or kinetic energy (\cite{priest2007}). This latter distinction 
is crucial for plasma thruster purposes, as the thermal energy would then need to be further converted into kinetic energy by 
means of a magnetic or solid nozzle.
In the previous work, the authors have determined that, in the configuration 
adopted for the analysis, during the first half stage of
the process  nearly half of the energy transferred to particles results to be
kinetic energy, becoming mainly thermal energy towards the end of the process.
The reader should bear in mind that within this work we are  considering a sole collisionless situation, 
meaning that any neutral-ion 
energy exchange is not taken into account here. On the other hand, an analysis of the effects with a partly-ionized plasma in the same configuration
could be an interesting 
subject for future works, even though it would 
require the use of specific collisional Particle-in-Cell codes.}
For those readers not entirely familiar 
with this process,
a visual representation of 
some typical magnetic reconnection outcomes is given in Fig. \ref{fig:0}.  These plots are obtained from a
computer simulation with the same setup as in  \cite{cazzola2016b}. On the lefthand side we show the out-of-plane magnetic
field generated from an initial in-plane $X-Y$ magnetic field profile. On the righthand side is shown the outcome in terms of
longitudinal component of the
ion velocity at the same time step.
 In \cite{cazzola2016b} upon performing a series of simulations set with typical laboratory plasma configurations, 
 the most promising outcome in terms of ion outflow velocity was further analyzed more into detail. 
The study revealed the great 
potentiality of magnetic reconnection to generate remarkable exhaust velocities and high specific impulse, achieving values
comparable with the presently existing plasma thruster technologies.
However, given the preliminarily nature of that analysis, important approximations were assumed, including:

\begin{itemize}
 \item simulations were performed in $2.5$ dimensions, neglecting the variation of vector quantities along the third direction 
 (namely the out-of-plane direction).
 This approximation may have led to different outcomes with respect to an actual $3D$ configuration.
 \item the analysis mainly focused on the energetics and the physical content in the outflow developed
 after reconnection under a particular initial configuration (i.e. an initially 
 not-equilibrium low-$\beta$ plasma). However, very little was mentioned concerning more applicative quantities, such as 
 the mass flow rate and, more importantly, the effective thrust possibly achievable out of such type of system, 
 which instead seem of greater importance for 
 potential laboratory applications.
 \item the ion exhaust analysis  was carried out over a single reconnection outflow wing. Despite the overall correctness of this
 choice, the almost longitudinal symmetric outflow generated by reconnection causes the ultimate net thrust gauge to be practically null.
 \item finally, all simulations were performed considering a plasma made out of Hydrogen. 
 On the other hand, it is now well established that, even though leading to remarkably higher exhaust velocities, this element shows
 unfortunate difficulties with the ionization process,
 so that other elements are instead preferred over it, such as Xenon, Hydrazine, \textcolor{black}{Argon and Krypton} (\cite{Mazouffre2016}). 
\end{itemize}

In this paper we therefore intend to address a more complete analysis of the process by relaxing most of the
approximations itemized earlier, such as:

\begin{itemize}
 \item a full $3D$ configuration is considered with the same parameters as in \cite{cazzola2016b}, 
 even though with a reduced mass ratio due to the significantly
 computational costs increase
 \item in line with more typical laboratory applications, we show results from $2.5D$ simulations set with a plasma made out of Xenon 
 at its first ionization state ($Xe^{+1}$)
 \item we aim to give new insights into the breaking of the outflow symmetry  necessary to obtain an effective net thrust out of this process. 
 Different solutions may be undertaken. The approach adopted here consists in setting up a density gradient along the longitudinal
direction (i.e. along $X$ in the reference used here), in order to attain an outflow unbalance between the two wings
 \item with regard to both the previous points, we intend to give more insights into the effective mass flow rates and thrust effectively 
 obtainable. This point is crucial for possible propulsion applications, considering that all currently existing plasma thrusters  
 hold the peculiarity of producing very high exhaust velocities against very low achievable thrusts, \textcolor{black}{this latter 
 mostly due to the limitations of 
 the electric energy suppliers on spacecraft, such as solar panels,}
 which makes this technology not particularly suitable for some specific applications, such as rapid spacecraft trajectory variations or gravity-escape maneuvers.
 
\end{itemize}

The paper is structured as follows. The upcoming Section \ref{sec:simulations} is devoted to explain the different setups adopted throughout the study. 
All the relevant results are shown in  Section \ref{sec:results}. Finally, Section \ref{sec:conclusions} is devoted to drawing some important conclusions.

\section{\label{sec:simulations} Simulations Setup}

All simulations have been performed by adapting 
the fully kinetic massively parallelized implicit moment particle-in-cell code iPIC3D (\cite{markidis2010}). 
This code has already been successfully used in a wide variety of different contents (e.g. \cite{deca2014,olshevsky2015b,innocenti2015,cazzola2016,broll2017}).
The code makes use of the following normalizations: lengths are normalized to the ion skin depth $d_i = \nicefrac{c}{\omega_{p,i}}$, where $\omega_{p,i} = \left( \nicefrac{4 \pi n_i e^2}{m_i} \right)^{0.5}$ is
the ion plasma frequency, as well as the time unit, velocities are normalized to the speed of light $c$, particle charges are 
normalized to the electric charge $e$ and 
masses are normalized to the ion mass $m_i$.
All the initial configurations consist in a single not-balanced current sheet domain within a $ \left( 20 \times 12 \right)\ \unit{d_i}$ box 
for the $2.5D$ simulations, and 
a $ \left( 20 \times 12 \times 8 \right)\ \unit{d_i}$ box for the  $3D$ setup.
A fixed Cartesian frame of reference is used with the $x$ coordinate being parallel to the current 
sheet, the $y$ coordinate being the direction of the magnetic field change, and the $z$ coordinate completing the reference set in the out-of-plane 
direction.
 As done in \cite{cazzola2016b},
a typical laboratory plasma density of 
$n = 10^{19}\ \unit{m^{-3}}$ is adopted and set uniform all over the domain. The ions and electrons temperature are fixed at, respectively, $T_i = 0.025\ \unit{eV}$ 
(i.e. room temperature), and $T_e = 10\ \unit{eV}$, 
while the magnetic field profile follows the traditional Harris 
profile \cite{harris1962,birn2001} normalized to the asymptotic field value of $B = 5000\ \unit{G}$, namely
\begin{equation}
 B_x \left(y \right) = B_0 \tanh \left( \frac{y}{\Delta} \right)
\end{equation}
with  $\Delta = 0.5$.
According to this density, the simulated box size results in a $\left( 144.2 \times 86.5 \right)\ \unit{cm}$ domain in $2.5D$, and 
$\left( 144.2 \times 86.5 \times 57.6 \right)\ \unit{cm}$ domain in $3D$
($d_i = 7.2\ \unit{cm}$). 
 As pointed out in \cite{cazzola2016b}, this specific initial setup partially recalls the not-force-free simulation studied in \cite{birn2009},
and particularly important is set a proper 
initial perturbation to trigger the reconnection process in a selected point, which lies in the middle for this case. 
Boundary conditions are chosen open along each boundary to better represent a realistic physical system. Particles and fields are free to escape the domain across the $x$ boundaries, 
with particles being re-injected from the $y$ borders, as similarly done in \cite{daughton2007} and \cite{wan2008}.
\textcolor{black}{Also notice that in the cases with Hydrogen plasma, both 2.5D and 3D, we are 
dealing with a fully ionized gas, whereas in the case with Xenon plasma we are dealing with an element ionized at its first ionization 
stage. On the other hand, all partcles are considered either fully or partly ionized, so that there is no production of neutrals.
A more detailed explanation on how a partly ionized plasma  has been handled in the simulations will be introduced in Sec. \ref{sec:Xe}.
}

\section{\label{sec:results} Results}

\subsection{Hydrogen Plasma - 2D} \label{sec:H2D}

\begin{figure}
 \centering
 \includegraphics[scale=.4]{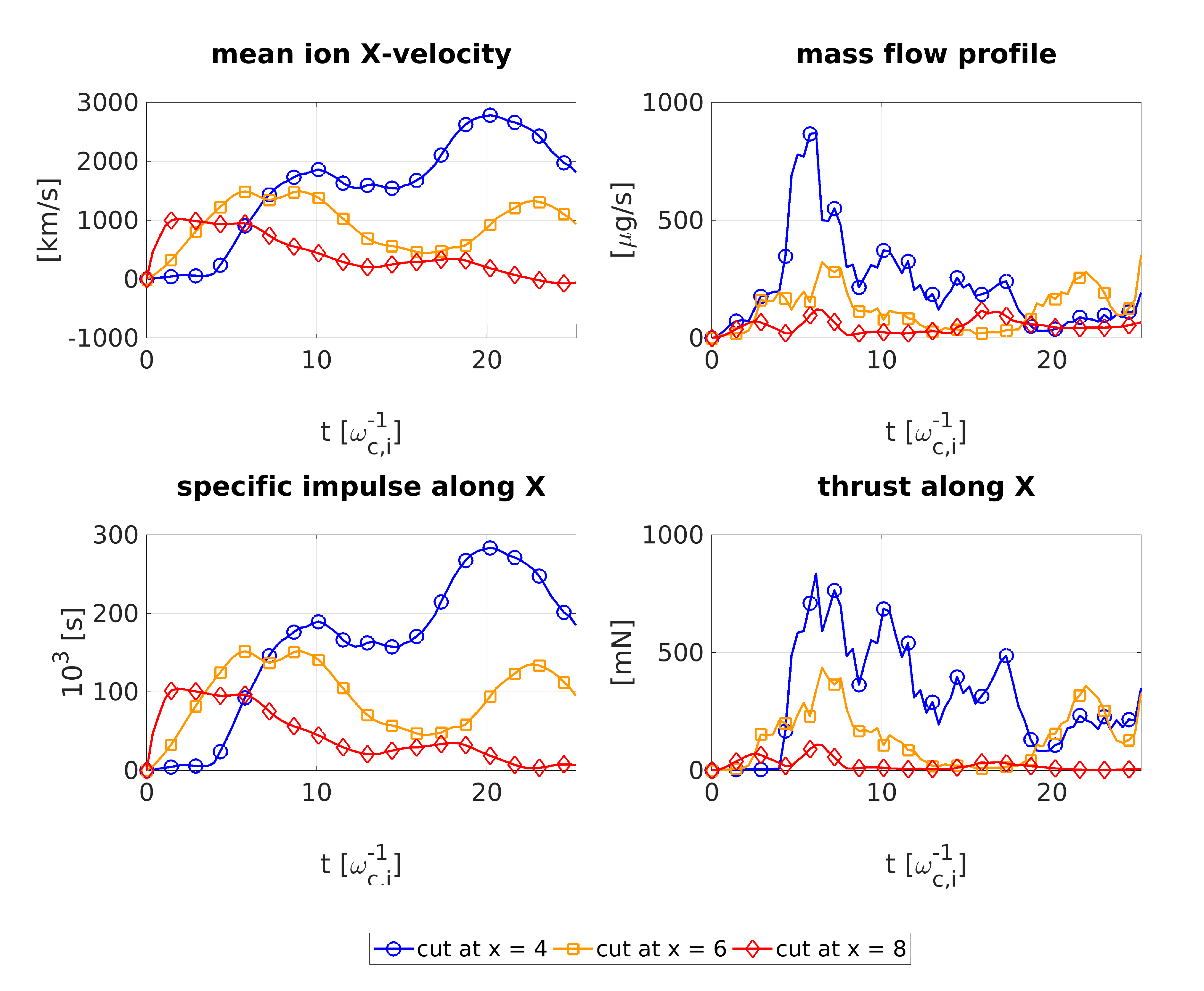}
 \caption{Four important quantities, namely the specific impulse, the mass flow rate and the thrust obtained out of Hydrogen plasma magnetic reconnection
 considering the only $X$-component of the ion velocity across the cross-sections at $x = 4$, $6$ and $8\ \unit{d_i}$. Plots have been slightly smoothed.}
 \label{fig:H2d1}
 \end{figure}

\begin{figure}
 \centering
 \includegraphics[scale=.4]{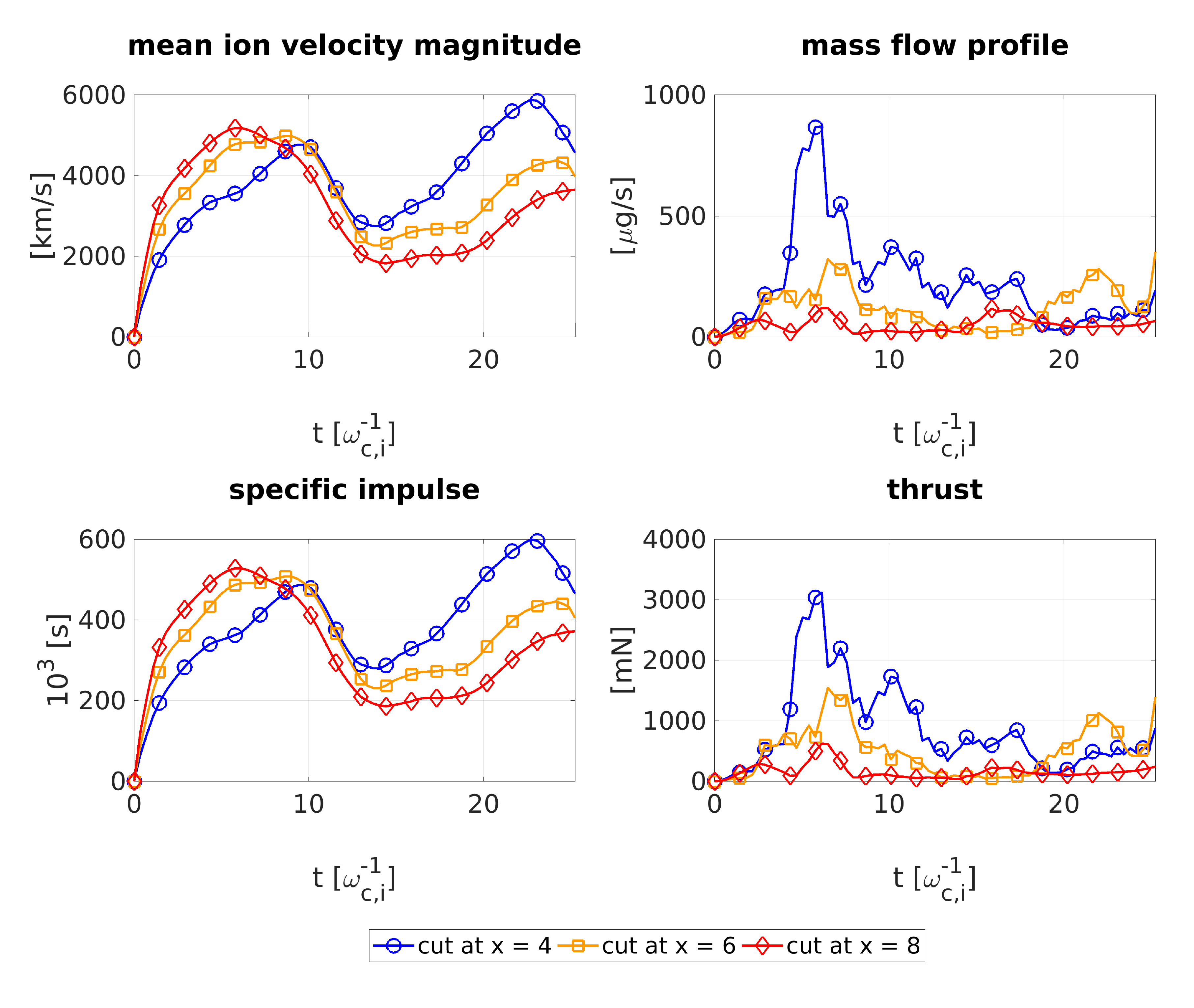}
\caption{Four important quantities, namely the specific impulse, the mass flow rate and the thrust obtained out of Hydrogen plasma magnetic reconnection
 considering the magnitude of the ion velocity across the cross-sections at $x = 4$, $6$ and $8\ \unit{d_i}$. Plots have been slightly smoothed.}
  \label{fig:H2d2}
 \end{figure}

In the work presented in \cite{cazzola2016b}, the authors mainly focused the analysis on
the physical content of this particular reconnection configuration, leaving momentarily out a more detailed analysis on other engineering quantities, such as 
the effective mass flow rate and, more importantly, the  thrust obtainable from this process. 
This section then intends to fill this gap by addressing an ad-hoc study on these important outputs under the same configuration.
Until now, the specific impulse has been consider as the most suitable parameter for a direct comparison with currently existing technologies.
This quantity is computed as the ratio between the exhaust velocity and the gravity constant. The key point is therefore what exhaust velocity to consider. 
In \cite{cazzola2016b}, this quantity was represented by  the average ion velocity along the $X$ direction, whose choice was mainly driven by 
the 2D nature of the analysis. However, the velocity components along $Y$ and $Z$ are also observed to be not negligible. 
We then propose here a study of the mass flow rate and the thrust
by considering two different approaches: on one hand we intend to consider again the sole $X$-component of the ion velocity 
as the main relevant longitudinal direction of an hypothetical
device, whilst on the other hand we intend to show the same outcomes as when the total average velocity magnitude is considered.
The mass flow rate is computed as
\begin{equation}
 \dot{m} = \left| \frac{m \left( t + \Delta t \right)  - m \left( t \right)  }{\Delta t} \right|  
\end{equation}
where $m$ is the outflow mass crossing a specific cross section. This mass is computed as the sum over all the masses along the specific cross section.  
Notice that $\dot{m}$ is being considered with its absolute value in order to give a better understand of the actual crossed mass. Being a scalar quantity, the 
sign would only indicate its direction upon setting up a reference direction.
From the simulation output, the mass in each cell is computed as 
\begin{equation}
 m = n_0 \cdot dx \cdot dy \cdot dz \cdot \left( m_i \rho_i + m_e \left| \rho_e \right| \right)
\end{equation}
where $\rho_{i,e}$ is the ions/electrons charge density, and $m_i$ and $m_e$ the respectively masses in $\unit{kg}$. $n_0$ is the initial particles density, which
is here set as $19\ \unit{m^{-3}}$.
By considering negligible any instantaneous velocity variation, as also done in the rocketry literature \cite{turner2008,Mazouffre2016},
we are able to compute the thrust from the momentum equation as
\begin{equation} \label{eq:flow}
 \mathbf{F} = \dot{m} \cdot \mathbf{V}.
\end{equation}
\textcolor{black}{Notice that for the case in study the mass flow rate $\dot{m}$ is determined only from the ions, 
as electrons have such a smaller mass that their total mass
can be considered negligible. Accordingly, the velocity in Eq. \ref{eq:flow} is the ions fluid velocity.}
Figures \ref{fig:H2d1} and \ref{fig:H2d2} show the results obtained for the 2.5D case with Hydrogen plasma. 
Notice that these plots have been slightly smoothed for a better readability.
The quantities are studied in the same cross sections as considered in \cite{cazzola2016b}, which may identify three different possible 
exit edges of the device. Besides the specific impulse,
three quantities are also represented, including the mean ion velocity, the mass flow rate expressed in $\unit{\nicefrac{kg}{s}}$
and the generated thrust expressed in $\unit{N}$.
Notice that the velocity is switched in sign for a better representation (i.e. multiplied by $-1$).
Unlike the case in \cite{cazzola2016b}, here the evolution is studied up to  the time of $25\ \unit{\omega_{c,i}^{-1}}$, as at later times 
the process can be considered finished. 
This reason will also be particularly helpful for the comparison with the cases shown later, 
which are overall seen to  finish earlier.

The specific impulse profile in Fig. \ref{fig:H2d1} pretty much resembles that shown in \cite{cazzola2016b}, with the only discrepancies
being most likely caused 
by the curve smoothing. In fact, in both the cases
we observe that the greatest value is reached across the outermost section at $x = 4\ \unit{d_i}$, whereby the reconnection
outflow is more developed, while  a lower magnitude peak is observed across each of the other two sections. In particular, a first
peak is observed across the nearest section  to the reconnection region, followed by a second increasing peak at nearly half-way of the 
outflow. Interestingly, across the farthest section we notice a double peaked temporal evolution, as if the outflow was being pushed out in two different stages, by ultimately
achieving very high velocity values. The latter can be explained by the presence of secondary reconnection processes occurring over the exhausts outflow.

The situation is seen drastically changed when the ion velocity  magnitude is instead considered. Here we observe the specific impulse to display
 as twice the values  as in the previous case, pointing out that the other two velocity components can be quite determinant. As in the previous case,
all the profiles show a double-peaked evolution, indicating a second acceleration occurring later in time.
The situation across the three cross-sections show quite a more uniform profile, with the peaks
being increasingly sorted according to their distance from it, although
displaying a very little discrepancy among their maximum values. 

Interesting is the evolution of the mass flow rate. This quantity is not influenced by the type of velocity considered. So the two cases show exactly the 
same behavior. We observe that the maximum mass crossing is always obtained in the earliest stage of the process, with the dominant value observed 
across the outermost cross-section. The flow rate is maintained fairly steady during the bulk of the reconnection process 
(i.e. between $7$ and $17\ \unit{\omega_{c,i}^{-1}}$).

\begin{figure}
 \centering
 \includegraphics[scale=.4]{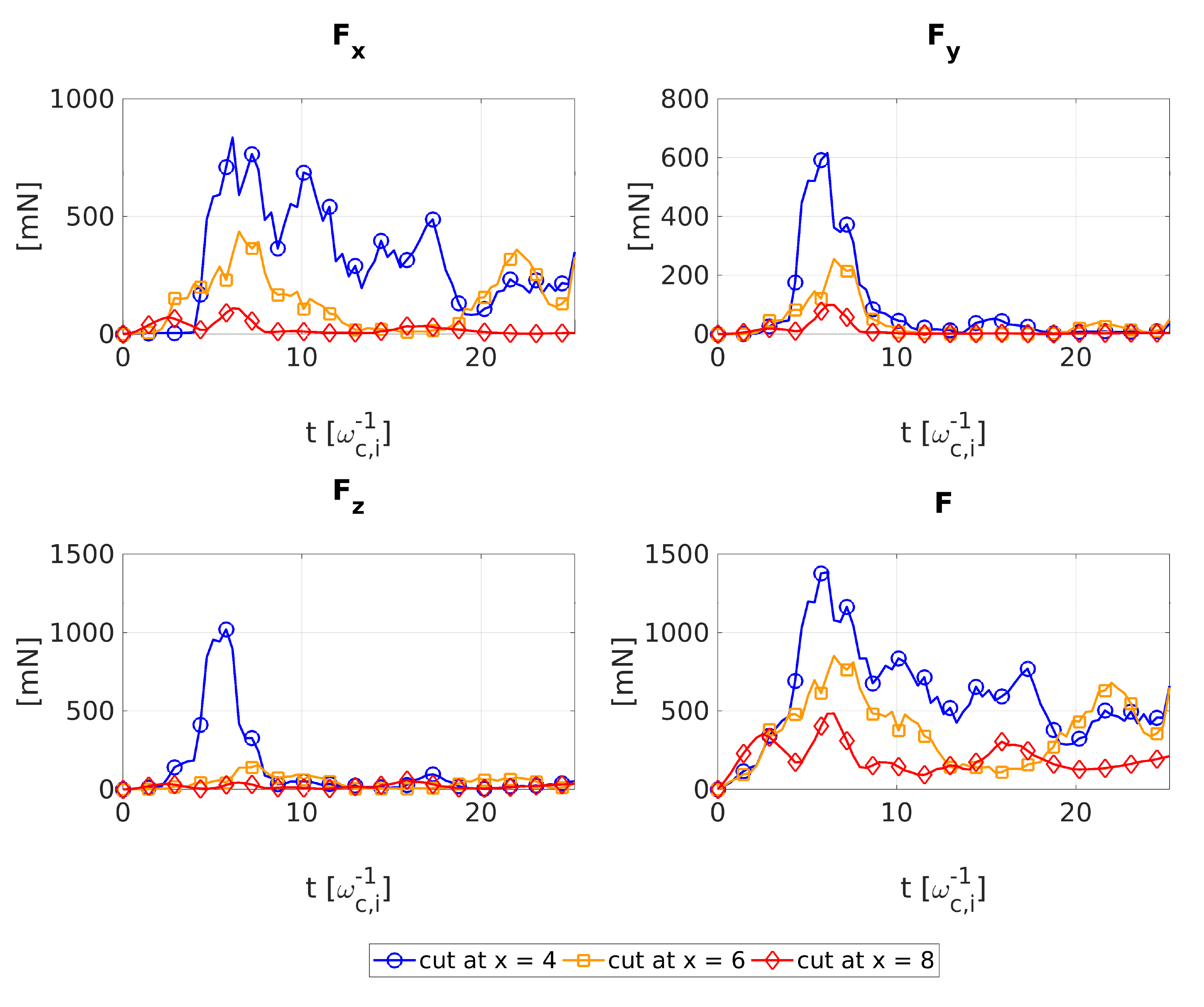}
 \caption{Evolution of the three thrust components out of magnetic reconnection for the case with Hydrogen plasma.}
 \label{fig:H2d3}
 \end{figure}

The thrust evolution follows the combination of both the mass flow rate and the velocity profile. 
We observe the highest thrust to be generated during the first stages of the process,
by maintaining after a sufficiently high outcome throughout most of the process. Overall, the most suitable outflow boundary for an 
hypothetical device results in being as far as 
possible from the reconnection region.
The thrust becomes more relevant as we consider the magnitude of the velocity rather than the single $X$ component. While the profile does not change
much with respect to the previous case, the magnitude instead makes itself much higher achieving values as high as $2 - 3\ \unit{N}$.
We therefore identified how the other thrust components may affect the overall evolution.
We plot them in Figure \ref{fig:H2d3}, where can be noticed that even though the $X$ direction  is still nearly 
the predominant force direction, the $Z$ and $Y$ components 
show to become the relevant ones for a very short time period in the region far out of the reconnection region. 
This result may need a further investigation in the future for 
the necessity  to have to counter-act these lateral forces and obtain a more balanced system. For instance, a solution could be adopting a symmetric double current
sheet configuration such that the overall lateral force balance result satisfied while maintaining the longitudinal force unaltered.

 \subsection{$\textbf{Xenon}^{+1}$ Plasma - 2D} \label{sec:Xe}

 As mentioned in the introduction, Hydrogen plasma is rarely employed for plasma propulsion purposes due to its difficulty in being easily ionized. 
 Upon this fact, other elements are currently considered, such as Xenon, Hydrazine, \textcolor{black}{Argon and Krypton} (\cite{turner2008, Mazouffre2016}), which show 
 lower ionization energies, yet
 at the cost of a much lower output exhaust velocity. It is in fact well-known that the achievable exhaust velocity 
 is inverse proportional to the atomic or molecular mass of the element used.
  In this section we show the results from an analysis considering a system similar to that shown in the previous section, 
  but a plasma made out of Xenon ionized to its 
 first ionization state. Even though the initial configuration is globally the same as that for the Hydrogen case,
 such a heavy element as Xenon$+1$
 requires us to sensibly lower the simulation mass ratio  down to $m_r = 512$ for a better 
 simulation performance (indeed, as iPIC3D is normalized to the ion mass, it is $m_r = m_{Xe} \cdot \frac{m_p}{m_e} = 67072$,
 with $m_{Xe} = 131\ \unit{uma}$ the Xenon atomic mass and $\frac{m_p}{m_e}$ the proton to electron mass ratio, here chosen equal to $512$).

\begin{figure}
 \centering
 \includegraphics[scale=.4]{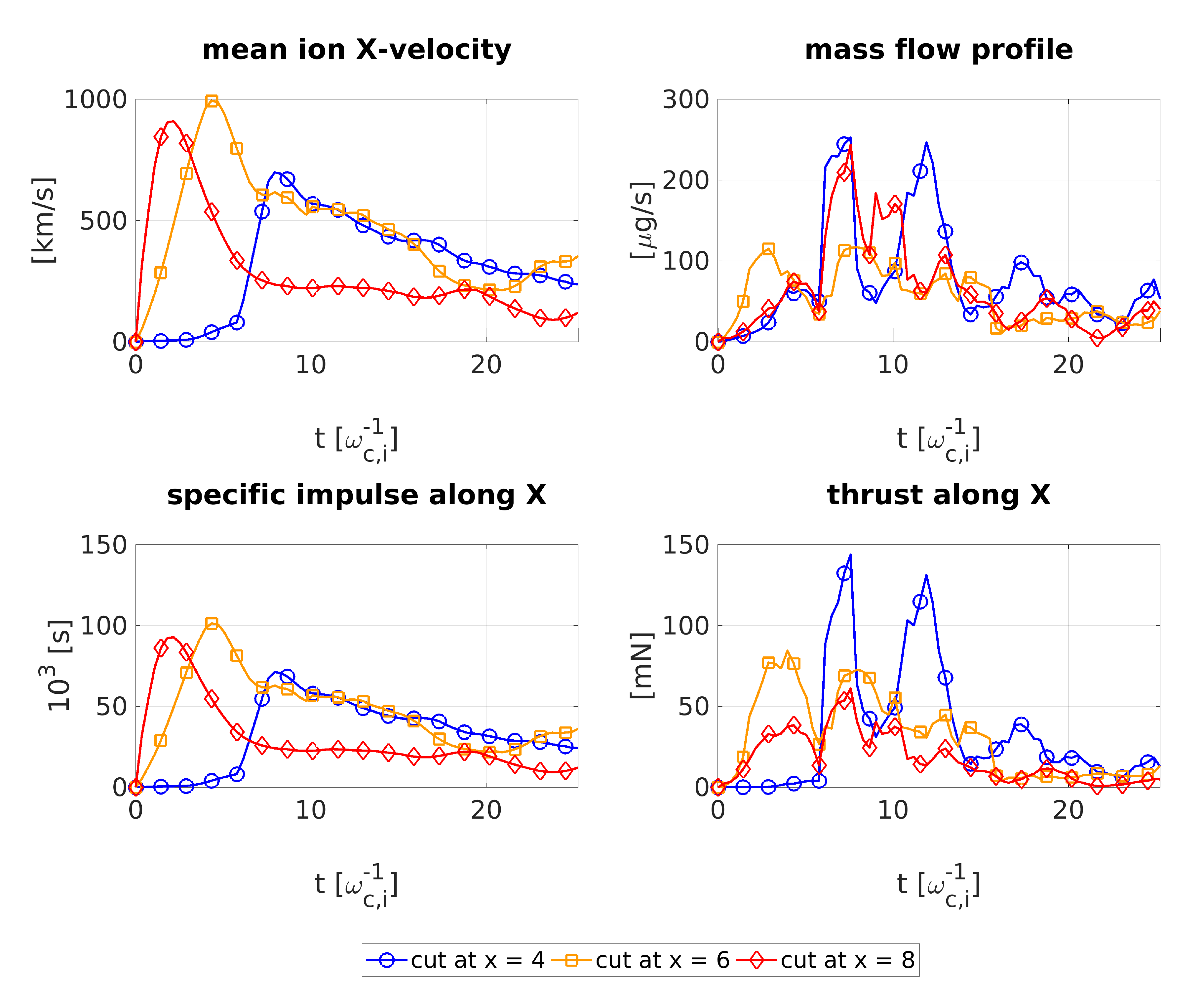}
 \caption{Temporal evolution of the specific impulse, mass flow rate and thrust obtained out of Xenon${+1}$ plasma magnetic reconnection
 considering the only $X$-component of the ion velocity across the cross-sections at $x = 4$, $6$ and $8\ \unit{d_i}$. Plots have been slightly smoothed. .}
 \label{fig:Xe2d1}
 \end{figure}

\begin{figure}
 \centering
 \includegraphics[scale=.4]{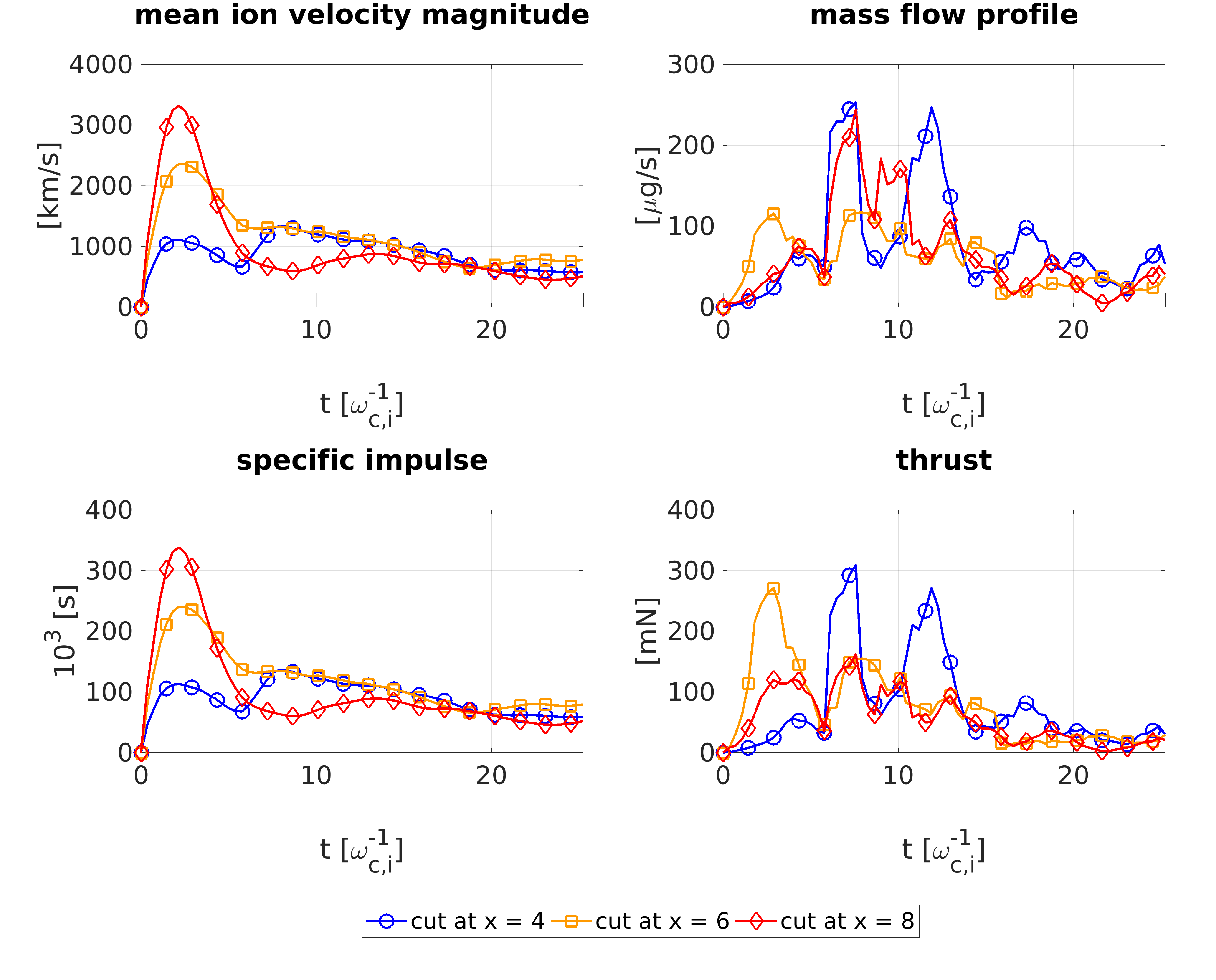}
 \caption{Temporal evolution of  the specific impulse, the mass flow rate and the thrust obtained out of Xenon${+1}$ plasma magnetic reconnection
 considering the magnitude of the ion velocity across the cross-sections at $x = 4$, $6$ and $8\ \unit{d_i}$. Plots have been slightly smoothed.}
 \label{fig:Xe2d2}
 \end{figure}

\begin{figure}
 \centering
 \includegraphics[scale=.4]{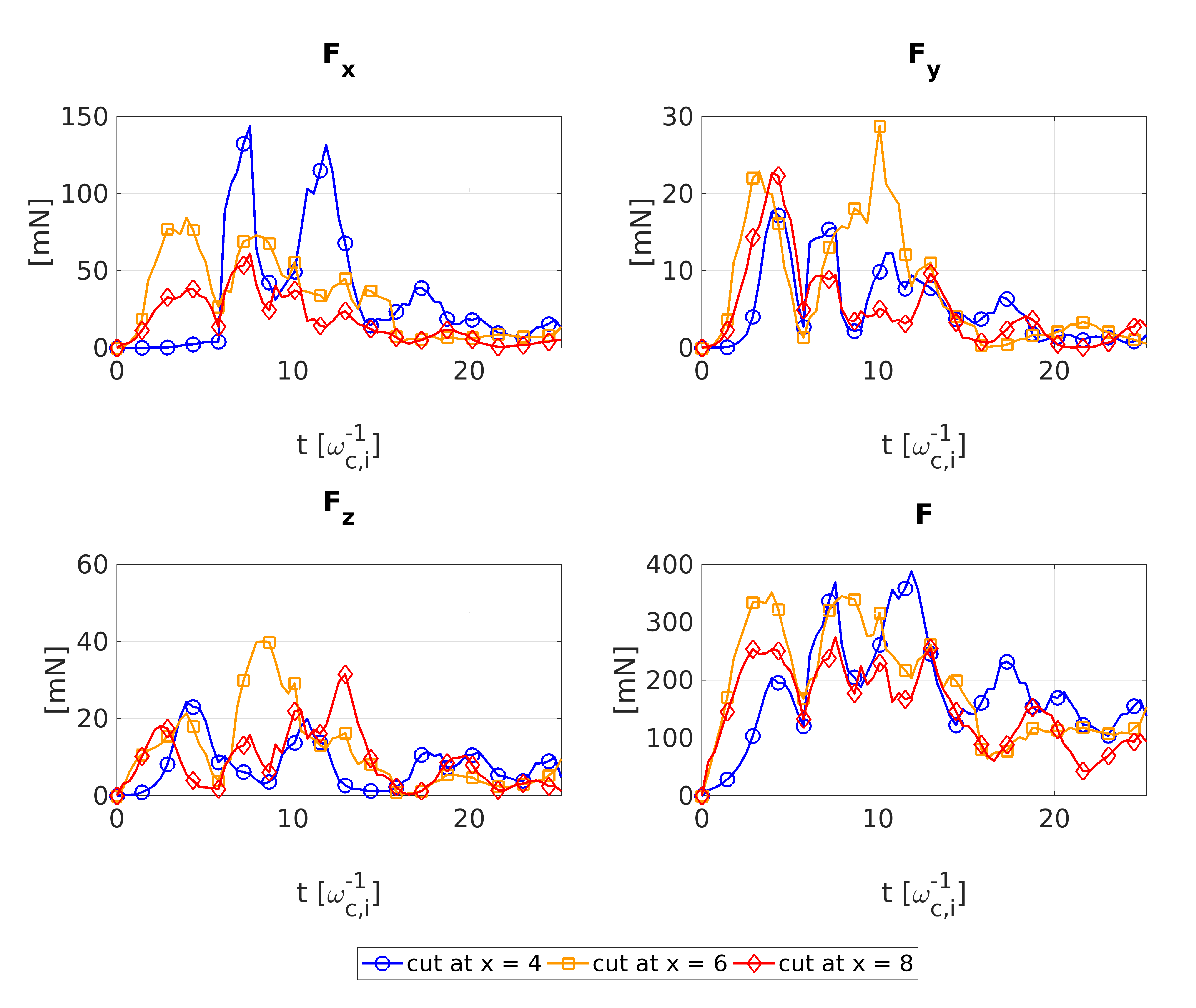}
 \caption{Temporal evolution of the three component of the thrust for the case with Xenon${+1}$ plasma.}
 \label{fig:Xe2d3}
 \end{figure}

 Figures \ref{fig:Xe2d1}, \ref{fig:Xe2d2} and \ref{fig:Xe2d3} show the same quantities shown earlier for the new case with Xenon.
 As expected, we observe a lower magnitude of the specific impulse due to the lower exhaust velocity caused by the Xenon's higher molecular mass. 
 However, unlike the previous case the situation along the $X$-component shows a series of peaks apparently not directly correlated to the distance from the reconnection region. 
 In particular, we notice that the maximum $I_{sp}$ value is reached in the middle of the outflow (i.e. section at $x = 6\ \unit{d_i}$).
 The latter suggests that magnetic reconnection in Xenon plasmas occurs faster than in Hydrogen plasmas, reaching out the top of its ion outflow velocity 
 at the very early stage of the process, yet still maintaining a steady outflow velocity profile. 
 Interesting is the mass flow profile, which shows the formation of a double-peaked profile shifting in time, with its highest magnitude 
 observed across the farthest cross-sections. The latter makes sense if we consider the flow tends to decelerate and accumulate towards the external boundary.
 Unlike the former case, however, we also observe that when the average magnitude of the velocity is considered, the situation results to be neater,
 with a lower $I_{sp}$ value than with the only single $X$ component. The latter suggests that the other two velocity components
 play a minor role. This is  also confirmed by the plots in Figure \ref{fig:Xe2d3}, which indicate that the thrust along the directions perpendicular 
 to $X$ are at least one order of magnitude lower. The thrust itself is however lower than the case with Hydrogen, most likely due to a lower ion exhaust 
 velocity achieved in this condition. The mass flow rate shows to be more or less in line with the previous case, even though with a much smoother 
 profile. Instead of the peaking observed across the farthest section, we now observe a more constant profile along all over the process, which might be
 a more preferable choice from a controllability point of view.
%

   \subsection{Symmetric Outflow Breaking}
   
\begin{figure}
 \centering
 \includegraphics[scale=.7]{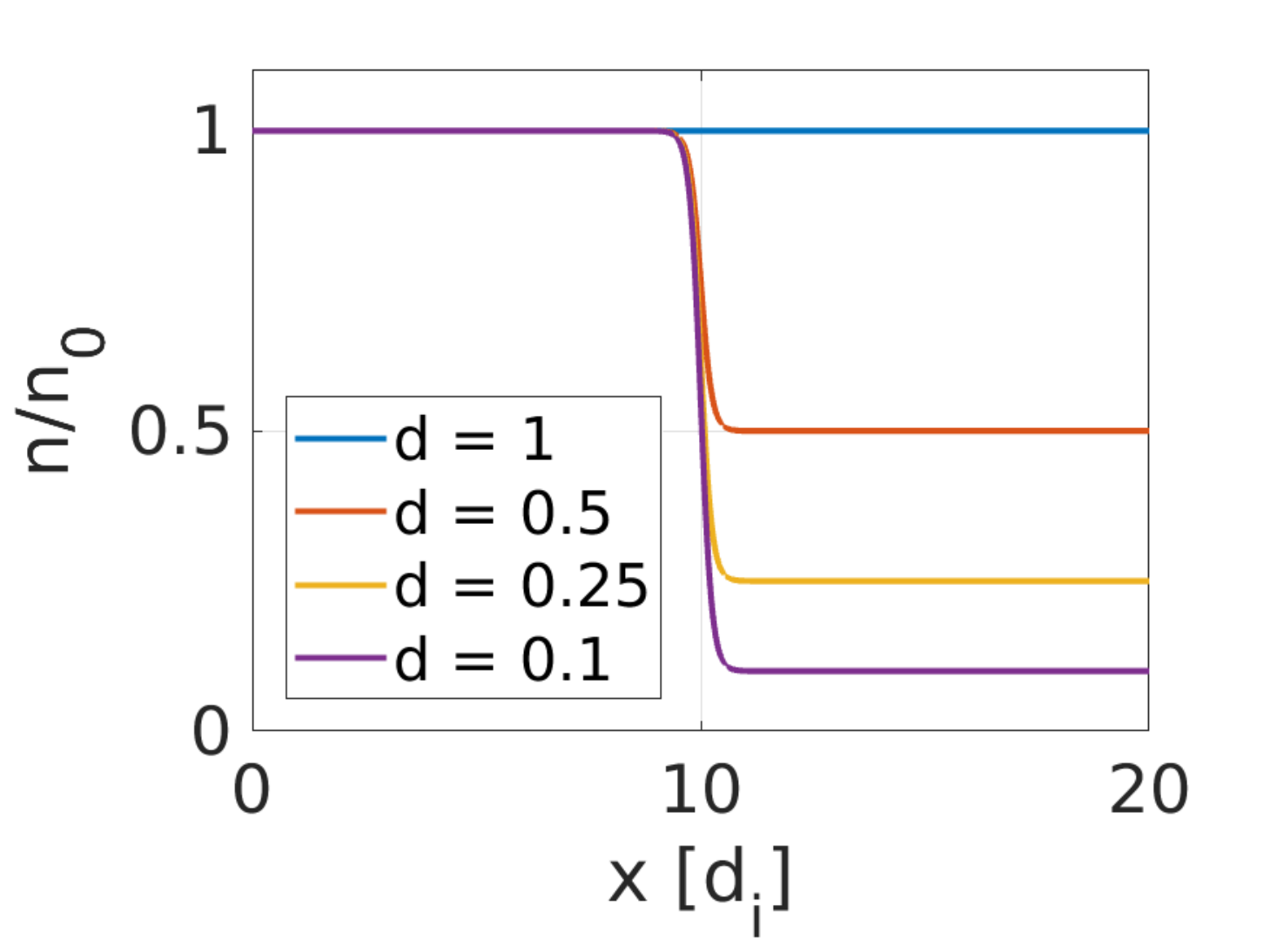}
 \caption{Initial density profiles adopted to break the left-right outflow symmetry and obtain a non-null net thrust gauge.}
 \label{fig:densH}
 \end{figure}

\begin{figure}
 \centering
 \includegraphics[scale=.4]{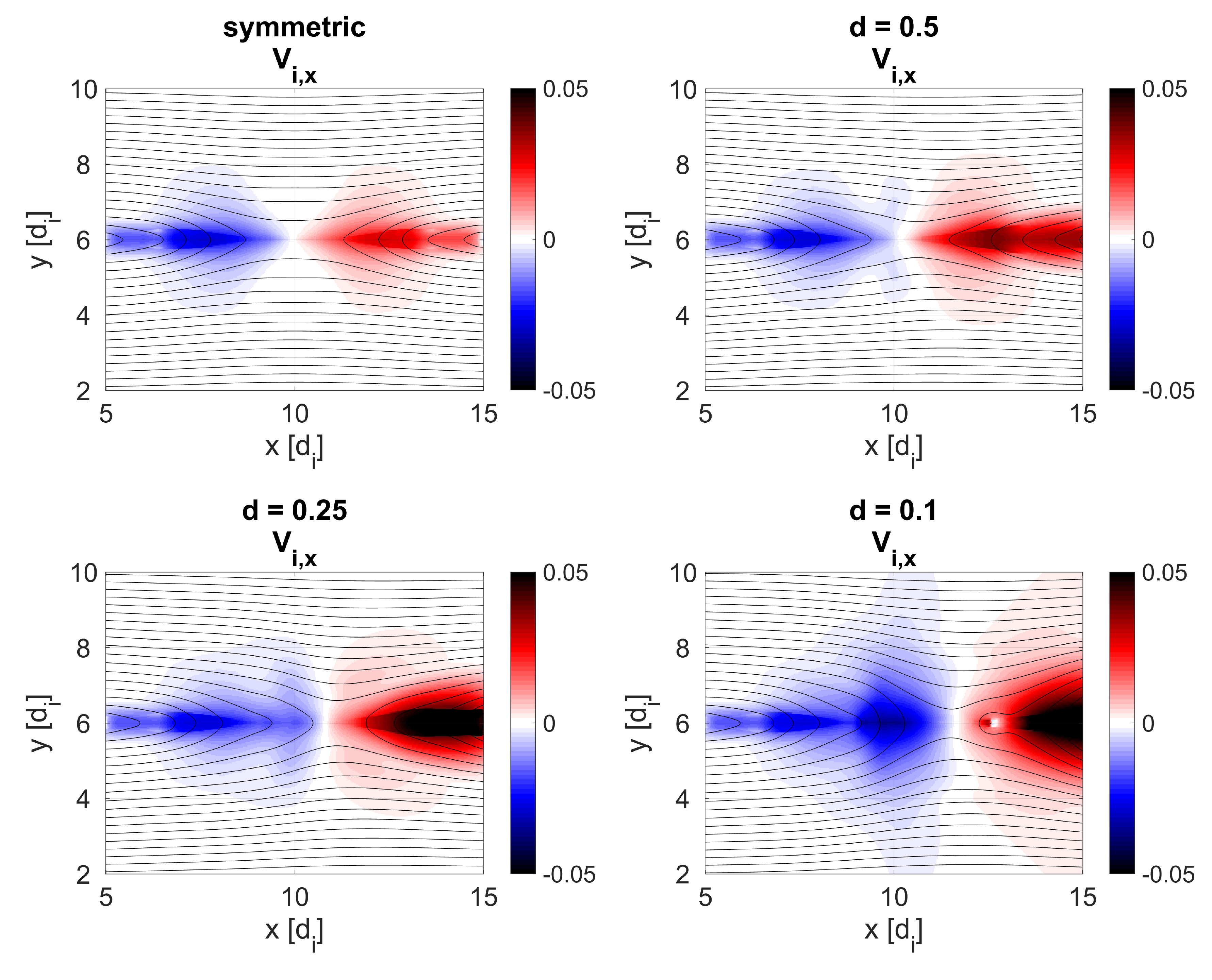}
 \caption{Comparison of the longitudinal ion velocity with different initial density profiles at the same time step.
 \textcolor{black}{Velocities are normalized to the light-speed.}}
 \label{fig:densH2}
 \end{figure}

\begin{figure}
 \centering
 \includegraphics[scale=.4]{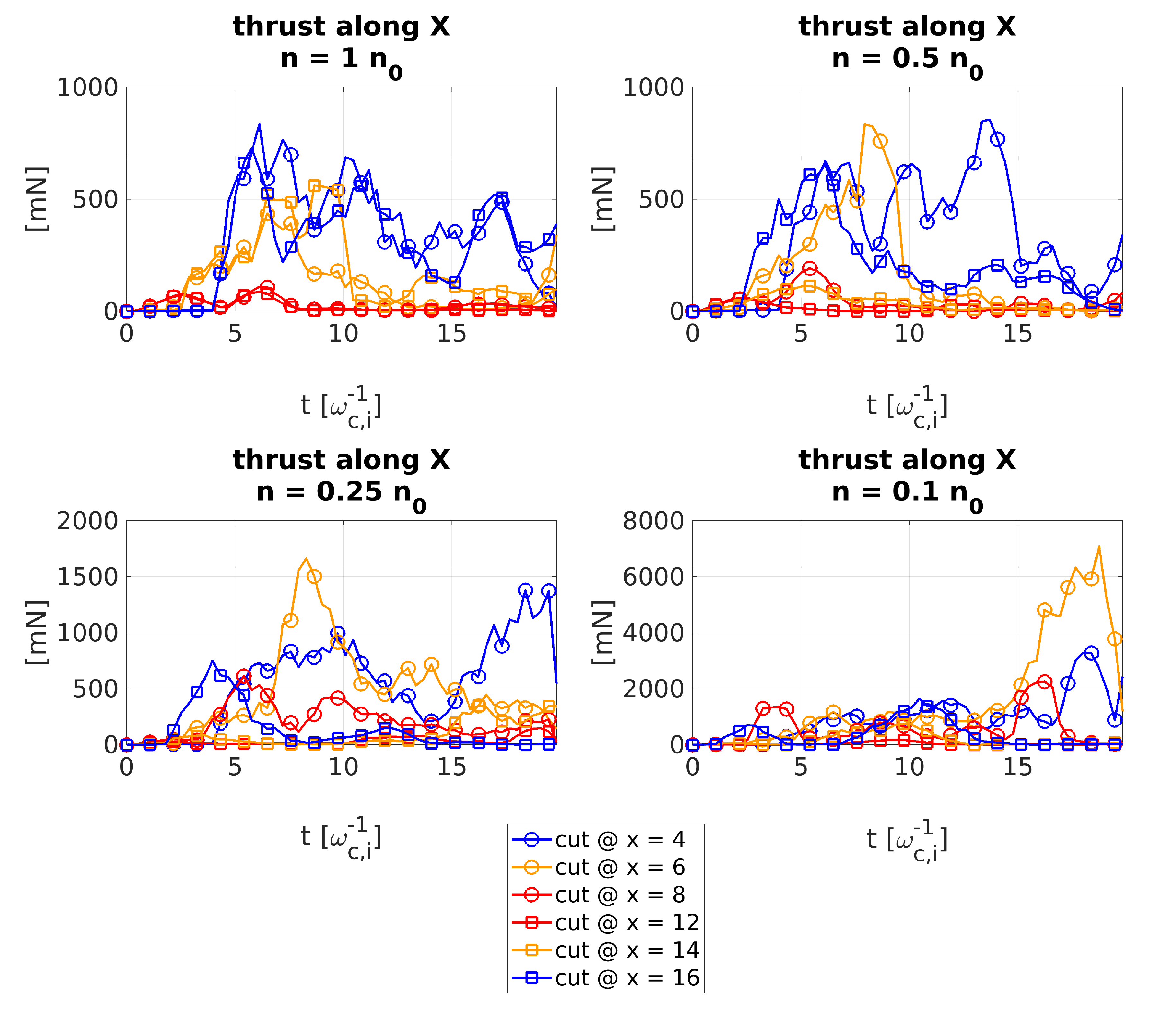}
 \caption{Temporal evolution of the thrust across the specularly located cross-sections in the two reconnection outflows when the only ion $X$-velocity component is
 considered. The color-code is kept uniform for each pair of
 corresponding specular cross-section to have a better comparison view.}
 \label{fig:densHdiffT}
 \end{figure}

\begin{figure}
 \centering
 \includegraphics[scale=.4]{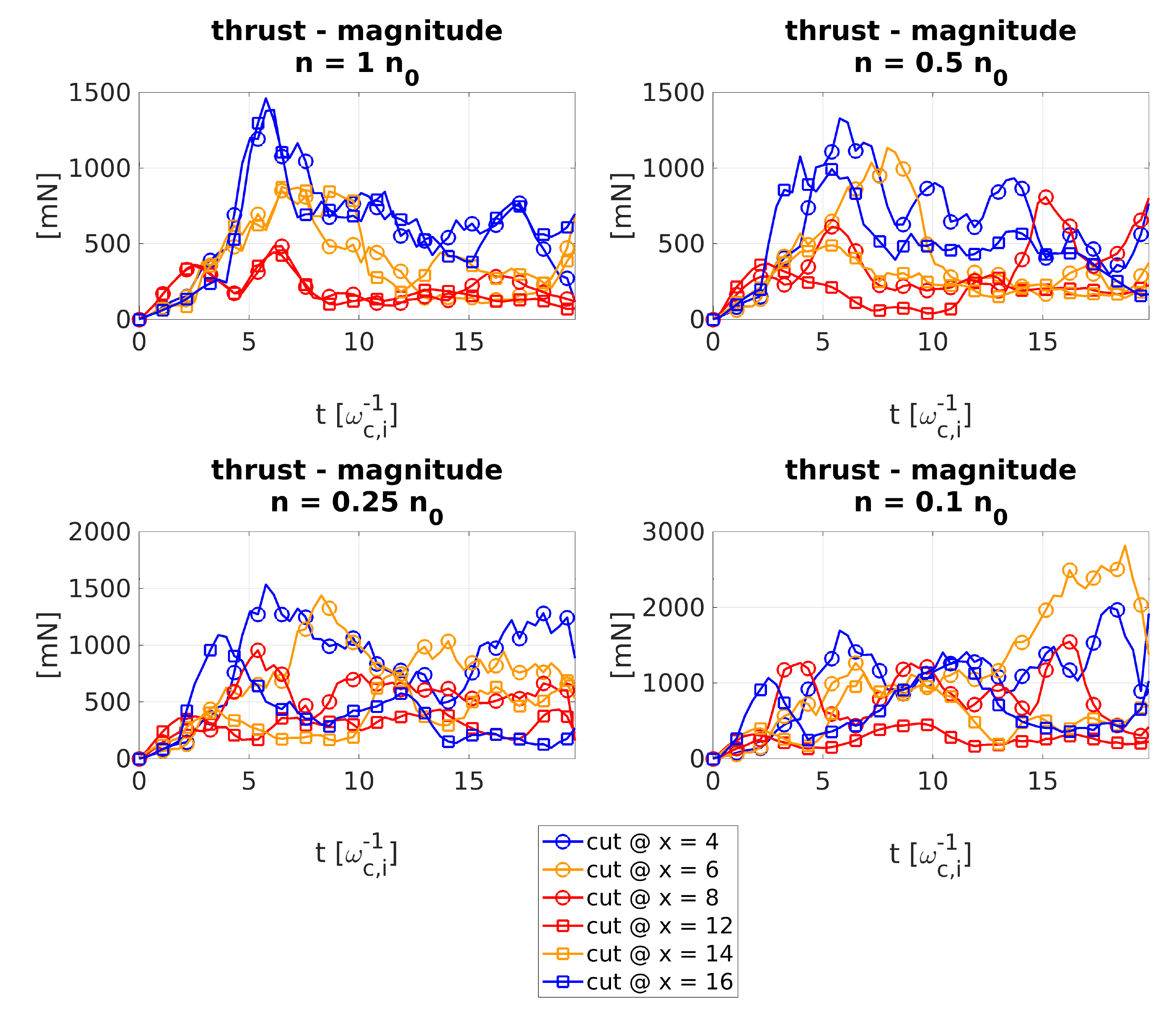}
 \caption{Temporal evolution of the thrust across the specularly located cross-sections in the two reconnection outflows when the magnitude of the 
 ion velocity is
 considered. The color-code is kept uniform for each pair of
 corresponding specular cross-section to have a better comparison view.}
 \label{fig:densHdiffTvi}
 \end{figure}

\begin{figure}
 \centering
 \includegraphics[scale=.4]{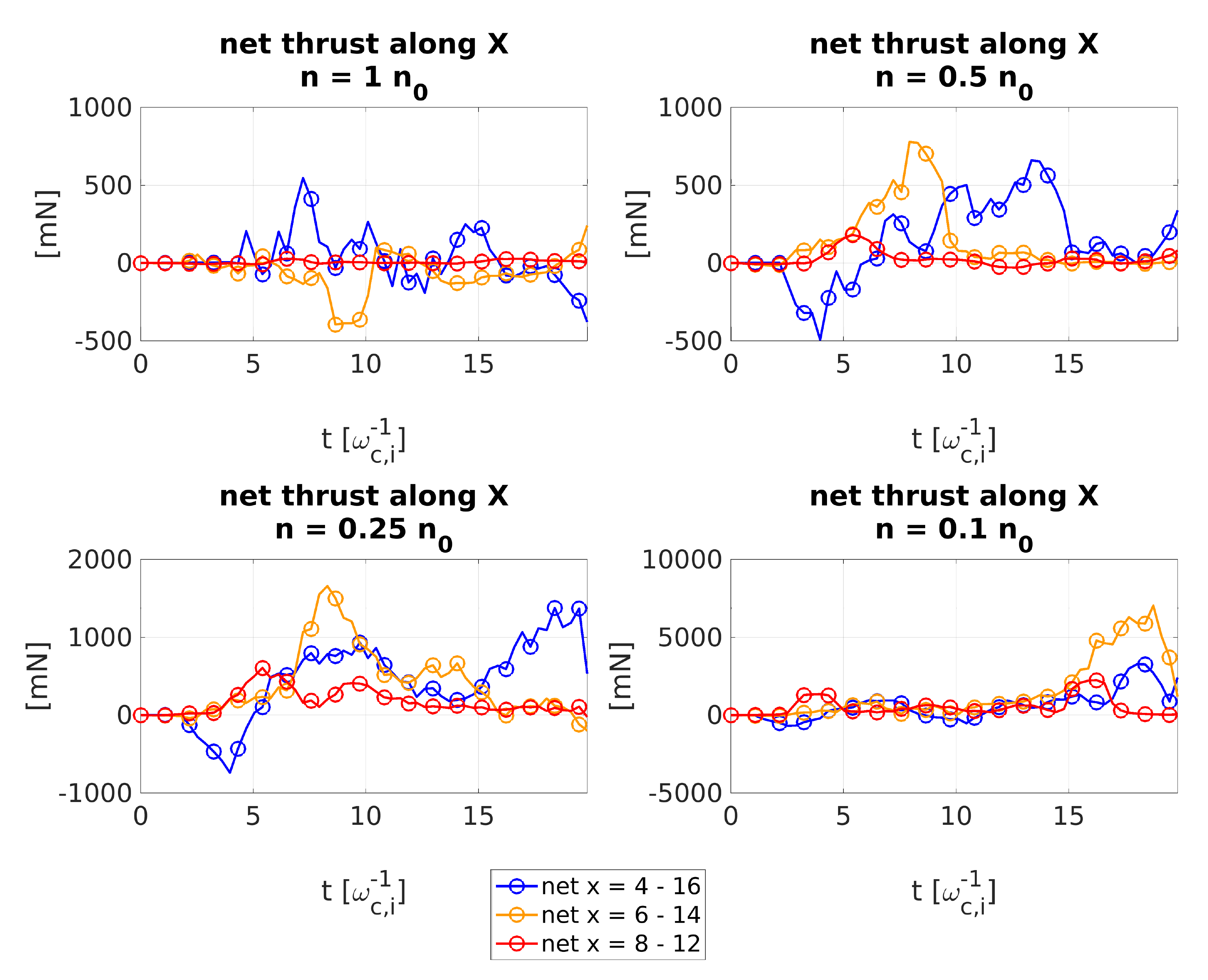}
 \caption{Net value of the thrust between the specular pair of cross-sections shown in Fig. \ref{fig:densHdiffT} when the only ion $X$-velocity component is
 considered. The plot shows the signed difference between the left-wing (asymptotic density) and the right-wind (density dropped).  The sign is ruled accordingly. }
 \label{fig:netdensHdiffT}
 \end{figure}

\begin{figure}
 \centering
 \includegraphics[scale=.4]{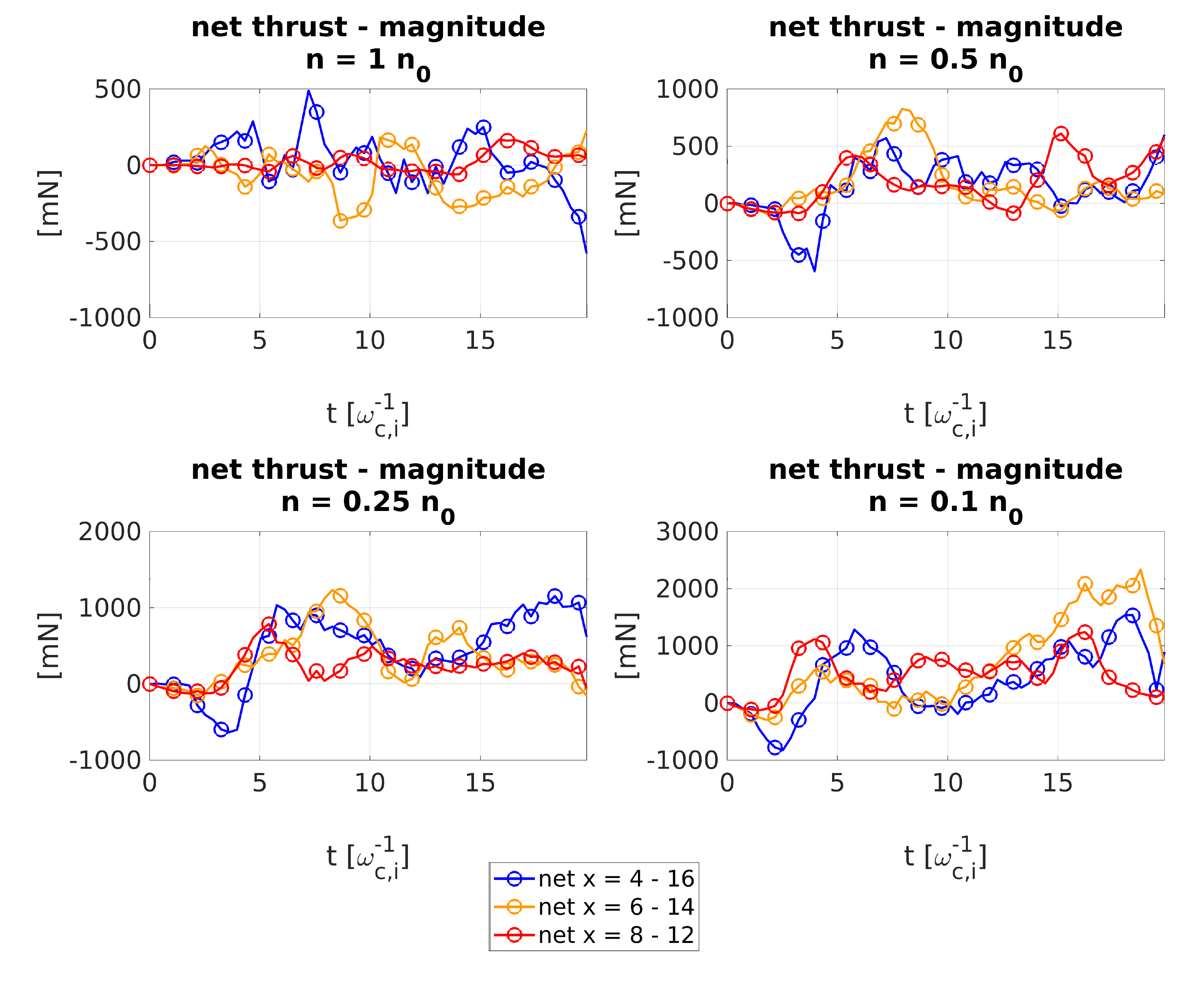}
 \caption{Net value of the thrust between the specular pair of cross-sections shown in Fig. \ref{fig:densHdiffT} when the magnitude of the ion velocity is
 considered. The plot shows the signed difference between the left-wing (asymptotic density) and the right-wind (density dropped).  The sign is ruled accordingly.}
 \label{fig:netdensHdiffTMAgn}
 \end{figure}

   As introduced earlier, one of the main point not addressed in \cite{cazzola2016b} was the fact that a nearly symmetric outflow 
   \textcolor{black}{expelled in two opposite directions cannot generate a net thrust,}
    causing the whole system  to stand in a perpetual steady position.
   This is also the reason why only one reconnection wing was indeed taken into consideration in that work. 
   In order to break this symmetry, several solutions may be undertaken.
   Here we propose  a solution based on a reconnection configuration resembling the situation in the outer the solar corona, and already considered 
   in earlier works (\cite{bettarini2010}).
   The underlying idea is to consider a longitudinal asymmetric density profile along the $X$ direction, by recalling 
   the different behavior hold by ions and electrons in entering the reconnection region (\cite{lapenta2014}). The magnetic field is instead still
   kept symmetric.
   A similar situation can also be appreciated from simulations of magnetic reconnection occurring at the magnetopause, whose transversal asymmetric profile 
   brings the reconnection outflows to take an asymmetric shape \cite{cazzola2015}. Given the different balance between the particles outflow and 
   the magnetic tension, the two wings now evolve differently from each other, giving out a ultimate different mass and velocity outflow, 
   therefore leading to an overall non-null net thrust.
      We then recalled the same simulation as explained in Section \ref{sec:H2D} but now featuring  a set of different density drops along the $X$ direction, 
      namely $50$, $25$ and $10\ \unit{\%}$ of the initial asymptotic value.
  The density profile now reads
  \begin{equation}
   n = 0.5 \cdot \left( n_0 + d \cdot n_0 \right) + \left( d \cdot n_0 - n_0 \right) \cdot \left[ \tanh \left( \frac{x}{\Delta_c} \right) \right]
  \end{equation}
  where $d$ is the decrease percentage, i.e. $d = 0.5,\ 0.25,\ 0,1$, and $\Delta_c$ indicates the drop steepness,
  set here as $\Delta_c = 0.25$. Different values may be used in future as a regulation parameter. Fig. \ref{fig:densH} shows the density profiles 
  here considered.
 Before analysing the mass flow rate, we give an overview of the different longitudinal velocity outcomes obtained with 
 these different density 
 profiles in Fig. \ref{fig:densH2}, where the asymmetric outflows can be neatly observed.
 We can now make a comparison of the thrust evolution between the same cross-sections considered earlier 
 and those specularly located in the other reconnection outflow, shown in Figs. \ref{fig:densHdiffT} and \ref{fig:densHdiffTvi}.
We intentionally kept the same color-code for each corresponding pair of specular sections to have an immediate view of the departure from each other.
To prevent from including in the analysis the formation of too many secondary islands, we preferred to interrupt the simulations at 
nearly $20\ \unit{\omega_{c,i}}$, when we can still consider the reconnection process amply concluded.
  We focus on the case along $X$ first. We notice that no great discrepancy is observed when the density is kept uniform. The only divergence between the two sides 
 is seen in the outermost sections (i.e. blue lines) at later times, and through the intermediate sections at intermediate times. This difference is likely to be
 caused by the occurrence of secondary reconnection events, which make the situation slightly asymmetric.
 However, we globally observe an overall consistency between the two sides. Such consistency is more and more broken as the density drop ratio increases.
  For convenience, in Figs.  \ref{fig:netdensHdiffT} and  \ref{fig:netdensHdiffTMAgn} we also plot the thrust difference between the corresponding sections. 
 Notice that the difference is set as the left-hand side minus the right-hand side, with the sign
 obtained accordingly.
 When the density is kept constant, a couple of net thrust peaks are seen to occur. Again, this can be explained with the formation of smaller secondary islands, 
 which go to influence the overall local thrust balance. In fact, the latter is mainly seen farther out of the reconnection region, whereby the 
 earlier collapse of the current sheet makes the outflow-islands interaction more plausible.
 A more interesting situation is depicted when an asymmetric density profile is set. The most advantageous situation seems to be 
 attained in the last case with one wing holding only $10\ \unit{\%}$ of the initial density.
 Here the net thrust achieves very high values at late times. However, such situation can be considered a limit case due to the onset of several 
 instabilities caused by such steep density gradient.
 The other two intermediate solutions underline that a density of a quarter of the initial value overall gives the best results, 
 despite the negative values shown at the early stage soon counteract with
 a global relevant gain over the rest of the evolution.  
 When the entire thrust is considered, we observe the situation to be much neater, with the case with density down to $10\ \unit{\%}$ showing 
 the most interesting outcomes. This indicates that 
 the other thrust components play a fundamental role in stabilizing the entire system.
 \textcolor{black}{Finally, we propose a summary of the time-average values of the net thrust for every cross-section pair in Table 
 \ref{table:1}. Even though the case with no density gradient does not show a null value as it should be expected,
 most likely due to the formation of secondary effects such as magnetic islands
 which may influence the local thrust in an asymmetric fashion, this value is so low that can certainly be considered
 negligible compared to the order of magnitude of the other cases (units are in $\unit{[mN]}$).}
\begin{center}\captionof{table}{Time-averaged evolution of the net thrust between the cross-section pairs to 
highlight the thrust gain as the set density gradient increases.}
\resizebox{\textwidth}{!}{\begin{tabular}[h!]{ c|c|c|c|c } 
\hline
Case & cross-section 4 - 16 [mN] & cross-section 6 - 14 [mN] & cross-section 8 - 12 [mN] & $\sum$ [mN] \\ \hline
$1 n_0$ & 112.64	& -232.70 &	71.88 & -48.18 \\ \hline
$0.5 n_0$ & 338.46 &	541.50 &	262.27 & 1142.24 \\ \hline
$0.25 n_0$ & 1001.50 &	1167.67 &	579.86 & 2749.03 \\ \hline
$0.1 n_0$ & 1258.16 &	2709.06 &	1449.63 & 5416.84  \\
\hline
\end{tabular}\label{table:1}}
\end{center}

  \subsection{$\textbf{Hydrogen}$ Plasma - 3D} \label{sec:3D}
  
\begin{figure}
 \centering
 \includegraphics[scale=.2]{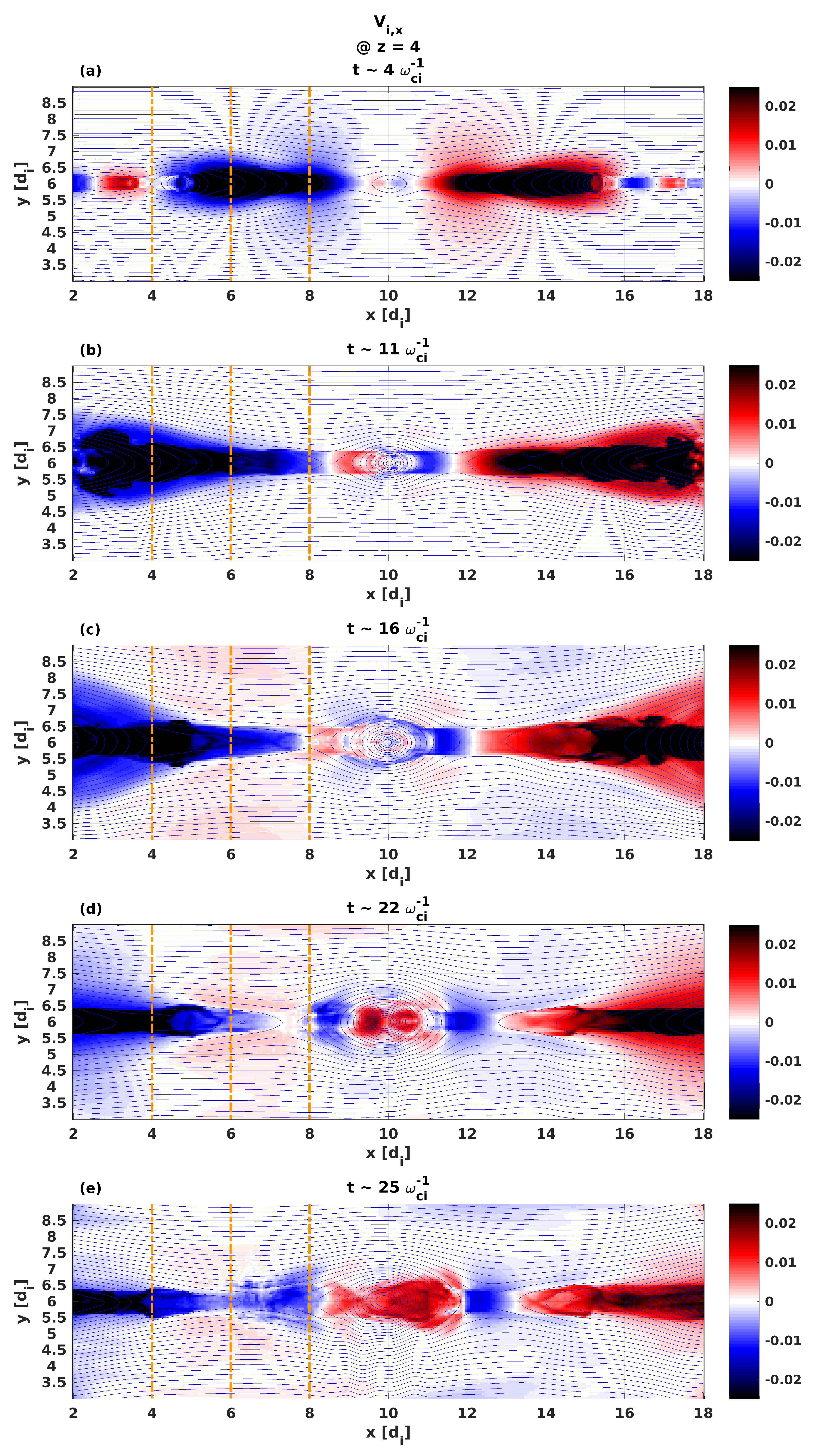}
 \caption{Profiles of the ions $X$-velocity component out of a 3D Hydrogen plasma magnetic reconnection simulation at $z = 4\ \unit{d_i}$.
 The yellow dashed lines represent the cross-section across which the analysis is performed.
 \textcolor{black}{Velocities are normalized to the light-speed.}}
 \label{fig:V1x3D}
 \end{figure}

\begin{figure}
 \centering
 \includegraphics[scale=.15]{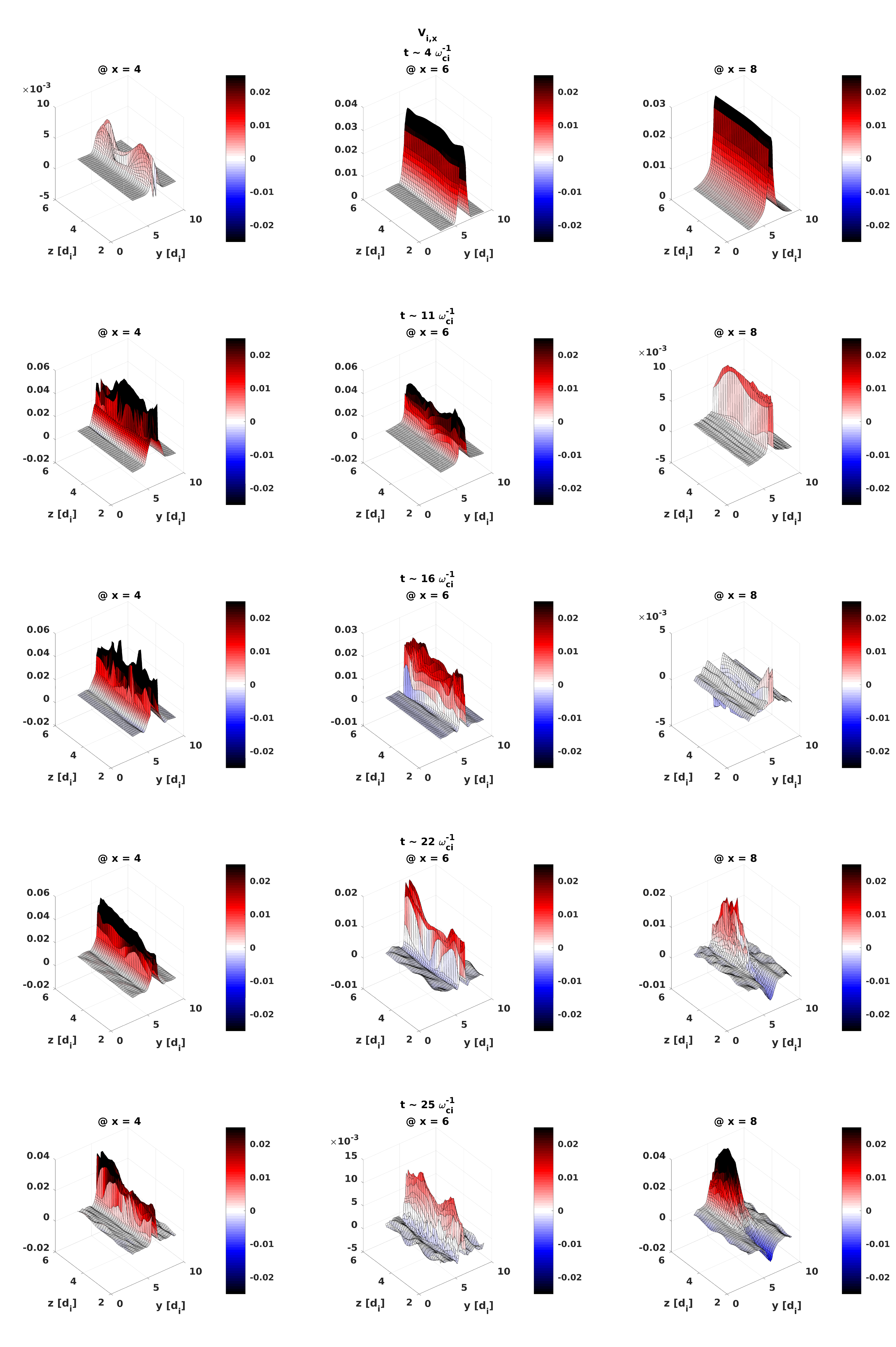}
 \caption{Surface representation of the ion $X$-velocity component across the cross-section indicated with yellow dashed lines in Fig \ref{fig:V1x3D}.
 \textcolor{black}{Velocities are normalized to the light-speed.}}
 \label{fig:cuts3D}
 \end{figure}

\begin{figure}
 \centering
 \includegraphics[scale=.4]{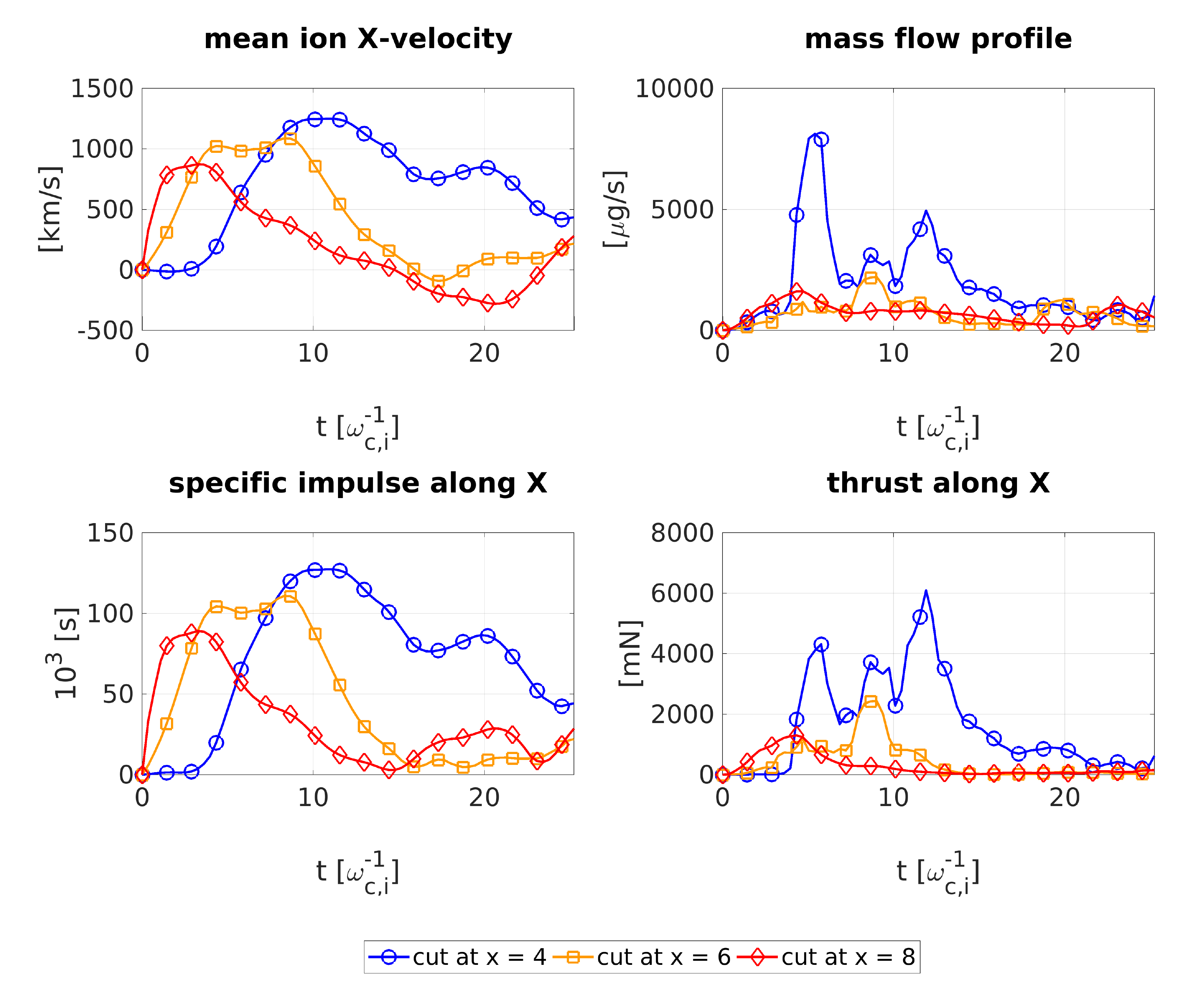}
 \caption{Temporal evolution across the same cross-sections of the specific impulse, mass flow rate and thrust when only the 
 $X$-component of the velocity is considered, from a 3D simulation of magnetic reconnection with Hydrogen plasma.}
 \label{fig:H3DVx}
 \end{figure}

\begin{figure}
 \centering
 \includegraphics[scale=.4]{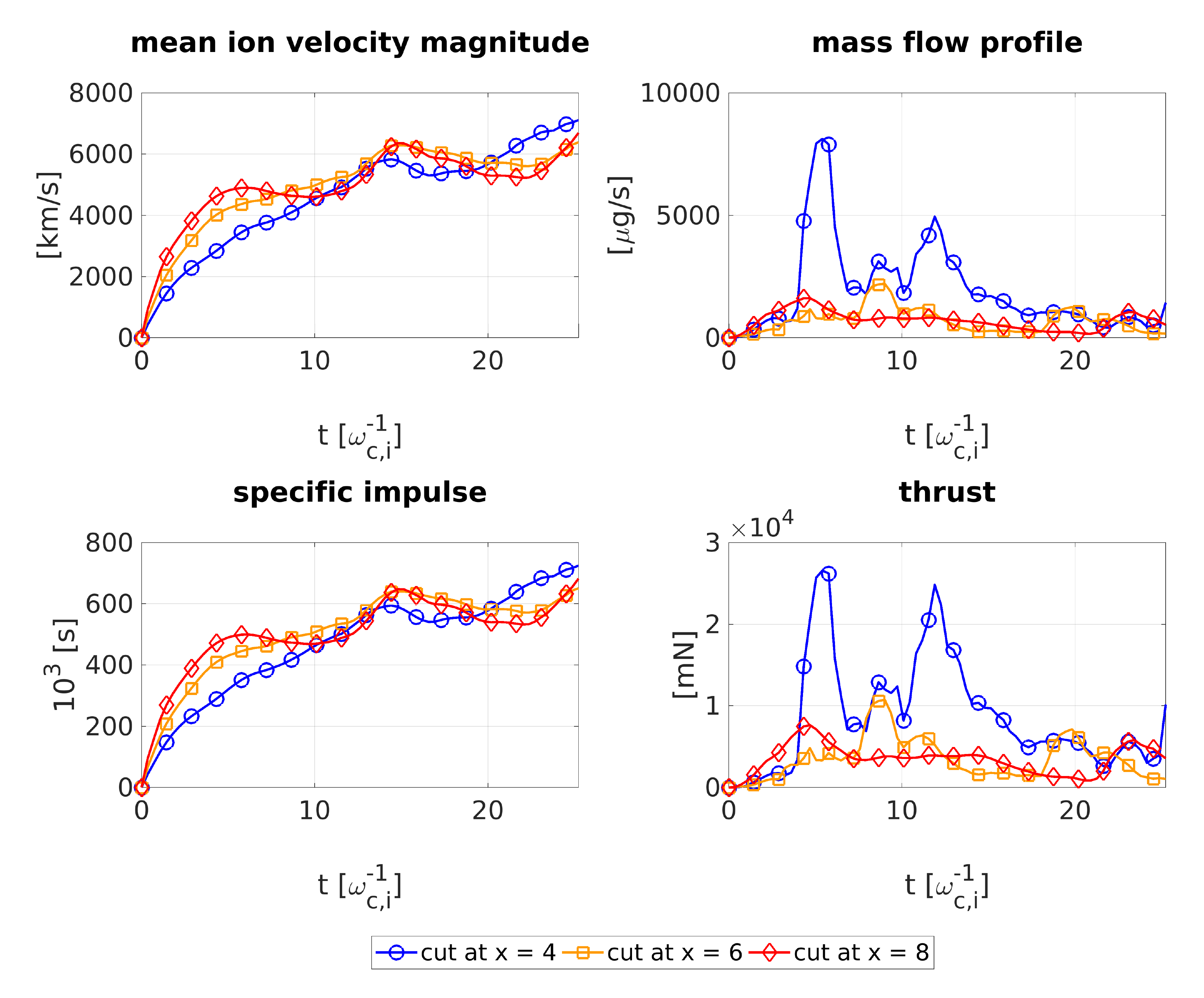}
 \caption{Temporal evolution across the same cross-sections of the specific impulse, mass flow rate and thrust when the magnitude
of the ion velocity is considered, from a 3D simulation of magnetic reconnection with Hydrogen plasma.}
 \label{fig:H3DVi}
 \end{figure}

\begin{figure}
 \centering
 \includegraphics[scale=.4]{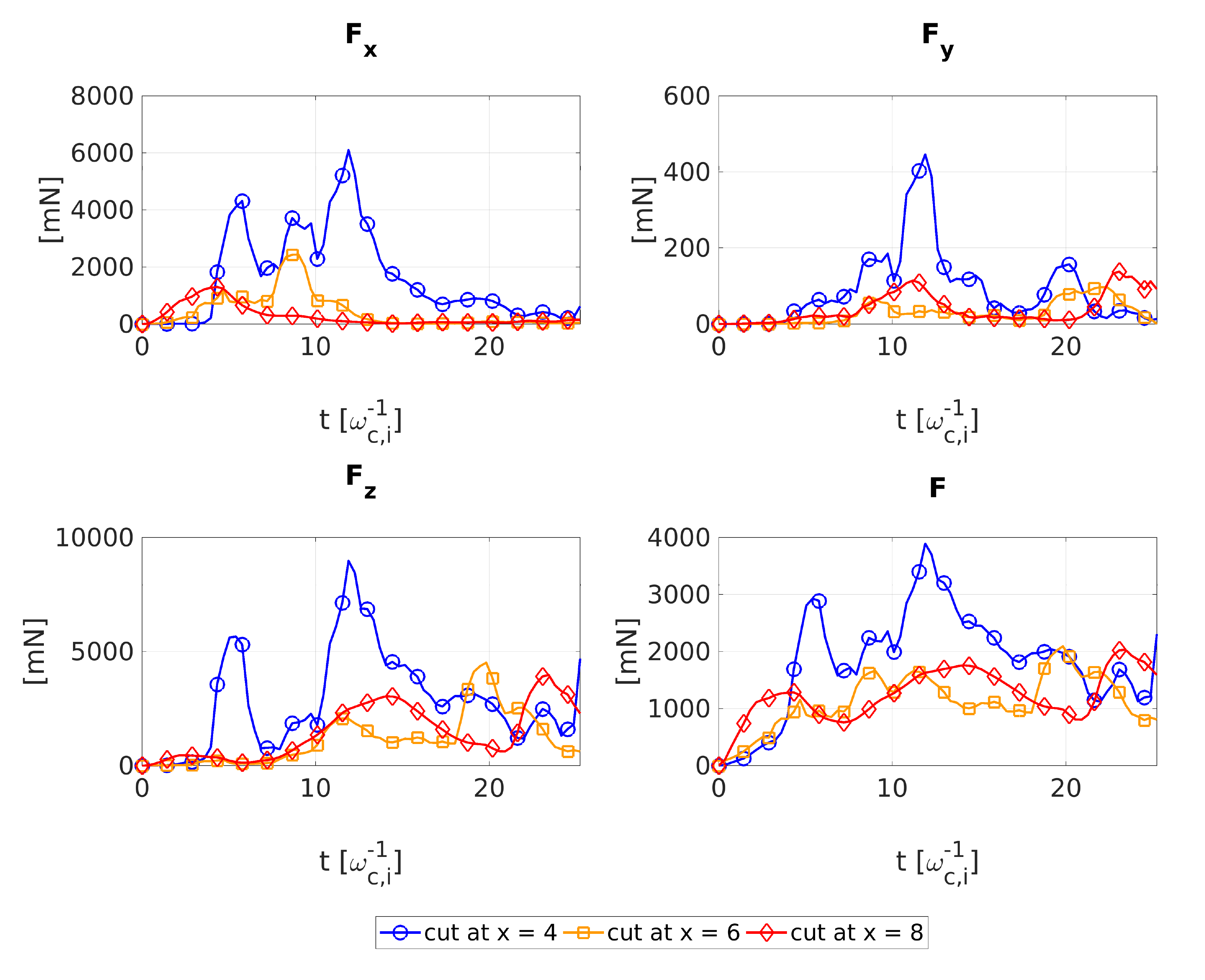}
 \caption{Temporal evolution of the thrust components out of a 3D simulation of magnetic reconnection with Hydrogen plasma. }
 \label{fig:H3DFx}
 \end{figure}

  All the analyses proposed so far were performed in 2.5D, with the third out-of-plane variation considered negligible. However, for a 
  more detailed analysis a full three-dimensional evolution would need to be 
  considered. In this Section we then propose a 3D evolution of the same configuration as in Section \ref{sec:H2D}.
  As introduced earlier, the third dimension (i.e. $Z$
  according to our frame of reference) is set as $8\ \unit{d_i}$, which translates into an overall domain-box of $\left( 144.2 \times 86.5 \times 57.6 \right)\ \unit{cm}$.
  By following the same approach as in the previous Sections, in Fig. \ref{fig:V1x3D} we give a first insight to the velocity evolution on 
  the $X-Y$ plane considering a cross-section through the central plane at $Z = 4\ \unit{d_i}$.
  Unlike the 2.5D case, we firstly notice that the outflow velocity is sensibly reduced (see \cite{cazzola2016b}). 
  A better outflow X-velocity perspective is given in 
   Fig. \ref{fig:cuts3D}, 
  where the surface evolution perpendicular to the cross-sections indicated with yellow dash lines in Fig. \ref{fig:V1x3D} 
  is plotted. Notice that, unlike the previous Figure,
  the sign of the velocity has been  multiplied by $-1$ 
  for the sake of a better representation. 
  The maximum outflow velocity is achieved in the early stage of the process, showing a regular and steady profile along the transversal direction
  (i.e. along $Z$). Also, notice the formation of the shock-shock
  discontinuities already observed in \cite{cazzola2016b}, which are visible at $t = 16\ \unit{w_{c,i}^1}$.
  The situation becomes more chaotic with time after the formation of new magnetic islands. The latter is particularly noticeable
  at $t = 25\ \unit{w_{c,i}}$, when
  the process is considered finished.
  We want to study the performance of the process in terms of mass flow rate and thrust across the same cross-sections considered earlier.
 Again, we distinguish between the case with the only $X$ outflow (Fig. \ref{fig:H3DVx}) and the case with the total ion velocity magnitude $V_i$ 
 (Fig. \ref{fig:H3DVi}).
 The plots represent the average velocity over the $Y-Z$ cross-sections taken into account.
 Fig \ref{fig:H3DVx} shows that the maximum mass flow is overall reached in the outermost section, similarly to the previous cases.
 The profiles show
 a three-peaked impulsive behavior only throughout this cross-section, whilst
 the other two sections show a more smoothed and regular behavior. 
 A similar evolution is also seen for the $X$ component of the total thrust. The maximum values are again reached in the outermost cross-section, 
 although 
 the maximum peak is now attained later in time.
 When the magnitude of the velocity is instead considered, the specific impulse shows greater values, with a similar thrust profile to that for the only 
 $X$ component, even though the values result now much higher. On the other hand, the velocity and specific impulse show a monotonically increasing behavior, 
 unlike the previous case, which showed more like 
 a \emph{shut down} evolution. Now the velocity, and the specific impulse accordingly, are seen to increase indefinitely.
 However, the same behavior is not seen for the thrust and the mass flow rate, which instead 
 \textcolor{black}{show a more fluctuating profile.}
 \textcolor{black}{We explain this behavior with the formation of secondary reconnection events, 
 which contribute in accelerating the 
 outflow. This latter
 is proved by the series of peaks observed in Fig. \ref{fig:H3DFx}, which gives an insight into each component of the outcome thrust.
 By comparing this Figure with Fig. \ref{fig:H2d3} we can notice the generation of multiple reconnection \emph{boosts} in the $Z$-direction.
 The phenomenon of secondary reconnection events in the outflow and the steady-state evolution of reconnection 
 are well explained in \cite{lapenta2015} and \cite{wan2008c}.}
 Finally, the $X$ and $Z$ components of the thrust appear to be dominant over the third direction, with the 
 out-of-plane thrust being slightly greater than that along the axis. Again, a possible technological solution to counteract this lateral thrust 
 might be the set of a symmetric double current sheet system, developing a thrust similar in magnitude 
 but opposite in direction.

\section{\label{sec:conclusions} Conclusions}

This paper presented a series of fully kinetic simulations of the process of magnetic reconnection to further demonstrate its potentiality for laboratory applications,
with special focus on possible applications in the plasma propulsion field.
For this purpose we used the fully kinetic massively parallelized implicit moment particle-in-cell code iPIC3D.
In the future more advanced updated versions of this scheme will eventually be used to reach even more precise results,
including the freshly proposed Energy Conserving scheme (\cite{lapenta2017,lapenta2017b}), 
as well as the application of the Quasi Neutral Limit approach to 
PIC schemes (\cite{tronci2015,camporeale2017,degond2017}), which allows to obtain neater simulations by getting rid of 
unwanted high-frequency waves.
In particular, following up what introduced in \cite{cazzola2016b}, we showed that, also when the approximations considered in that work are relaxed, a technology system based on 
this physical process
can still be considered highly competitive over the currently existing plasma propulsion technologies.
More into detail, we proposed here an analysis of important engineering quantities, including specific impulse, mass flow rate and outcome thrust
under different magnetic reconnection configurations, including a 3D simulation with 
with Hydrogen plasma, and 2.5D simulations with a plasma  from Xenon 
ionized to its first state. 
Also, given the non-negligible contribute brought up by the transversal and out-of-plane components  of the velocity (here pointed out as $Y$ and $Z$ respectively),
these quantities  have been studied in terms of both the sole $X$ component (axial component) and the total velocity magnitude. The two cases show remarkably 
different
results, which indicate that the perpendicular components of the velocity may have an important relevance in the final system design.
Finally, one of the main issues hindering the transfer of this technology from theory to a practical realization 
is the nearly symmetric outflow developed by magnetic reconnection,
which would inevitably lead to a null net thrust.
In order to break this symmetry without compromise the physical content of the process, here we have proposed 
the idea of considering an asymmetric longitudinal profile of the initial density. Based upon the  behavior undertaken by particles 
entering the reconnection region,
we have demonstrated that a remarkable net thrust between the two reconnection wings can be obtained by setting up
a significant density gradient over 
the axial direction. A 
lower limit threshold should yet still be considered to prevent from the raise of instabilities caused by such sharp mass difference.

%

\subsection*{Acknowledgments}

The present work has begun within the framework of the 
Illinois-KULeuven Faculty/PhD Candidate Exchange Program held in 2015.
E.C. acknowledges support from the Leverhulme Research Project Grant Ref. 2014-112 and would like to thank Dr. Cesare Tronci for the opportunity 
given to carry on this work.
Part of the work done by E.C. was performed over his permanence at the Centrum Wiskunde \& Informatica (CWI) in Amsterdam during a collaboration with Dr. Enrico Camporeale,
to whom the author is thankful for the enlightening discussions on the PIC methods.
The simulations were conducted on the computational resources provided by the PRACE Tier-0 machines (Curie, Marconi and MareNostrum III supercomputers) and
on the Flemish Supercomputing Center (VSC-VIC3).

 
   \bibliographystyle{plain}

\begin{thebibliography}{10}

\bibitem{bettarini2010}
Lapo Bettarini and Giovanni Lapenta.
\newblock Spontaneous non-steady magnetic reconnection within the solar
  environment.
\newblock {\em Astronomy \& Astrophysics}, 518:A57, 2010.

\bibitem{birn2001}
J~Birn, JF~Drake, MA~Shay, BN~Rogers, RE~Denton, M~Hesse, M~Kuznetsova, ZW~Ma,
  A~Bhattacharjee, A~Otto, and P.L. Pritchett.
\newblock Geospace environmental modeling (gem) magnetic reconnection
  challenge.
\newblock {\em Journal of Geophysical Research: Space Physics (1978--2012)},
  106(A3):3715--3719, 2001.

\bibitem{birn2009}
Joachim Birn and Michael Hesse.
\newblock Reconnection in substorms and solar flares: analogies and
  differences.
\newblock 27(3):1067--1078, 2009.

\bibitem{biskamp00}
D.~Biskamp.
\newblock {\em Magnetic Reconnection}.
\newblock Cambridge University Press, 2000.

\bibitem{broll2017}
JM~Broll, SA~Fuselier, and KJ~Trattner.
\newblock Locating dayside magnetopause reconnection with exhaust ion
  distributions.
\newblock {\em Journal of Geophysical Research: Space Physics}.

\bibitem{camporeale2017}
Enrico Camporeale and Cesare Tronci.
\newblock Electron inertia and quasi-neutrality in the weibel instability.
\newblock {\em Journal of Plasma Physics}, 83(3), 2017.

\bibitem{cazzola2015}
E~Cazzola, ME~Innocenti, S~Markidis, MV~Goldman, DL~Newman, and G~Lapenta.
\newblock On the electron dynamics during island coalescence in asymmetric
  magnetic reconnection.
\newblock {\em Physics of Plasmas (1994-present)}, 22(9):092901, 2015.

\bibitem{cazzola2016b}
Emanuele Cazzola, Davide Curreli, Stefano Markidis, and Giovanni Lapenta.
\newblock On the ions acceleration via collisionless magnetic reconnection in
  laboratory plasmas.
\newblock {\em Physics of Plasmas (1994-present)}, 23(11):112108, 2016.

\bibitem{cazzola2016}
Emanuele Cazzola, Maria~Elena Innocenti, Martin~V Goldman, David~L Newman,
  Stefano Markidis, and Giovanni Lapenta.
\newblock On the electron agyrotropy during rapid asymmetric magnetic island
  coalescence in presence of a guide field.
\newblock {\em Geophysical Research Letters}, 2016.

\bibitem{daughton2007}
William Daughton and Homa Karimabadi.
\newblock Collisionless magnetic reconnection in large-scale electron-positron
  plasmas.
\newblock {\em Physics of Plasmas (1994-present)}, 14(7):072303, 2007.

\bibitem{deca2014}
Jan Deca, Andrey Divin, Giovanni Lapenta, Bertrand Lemb{\`e}ge, Stefano
  Markidis, and Mih{\'a}ly Hor{\'a}nyi.
\newblock Electromagnetic particle-in-cell simulations of the solar wind
  interaction with lunar magnetic anomalies.
\newblock {\em Physical review letters}, 112(15):151102, 2014.

\bibitem{degond2017}
Pierre Degond, Fabrice Deluzet, and David Doyen.
\newblock Asymptotic-preserving particle-in-cell methods for the
  vlasov--maxwell system in the quasi-neutral limit.
\newblock {\em Journal of Computational Physics}, 330:467--492, 2017.

\bibitem{harris1962}
Eo~G Harris.
\newblock On a plasma sheath separating regions of oppositely directed magnetic
  field.
\newblock {\em Il Nuovo Cimento Series 10}, 23(1):115--121, 1962.

\bibitem{innocenti2015}
ME~Innocenti, M~Goldman, D~Newman, S~Markidis, and G~Lapenta.
\newblock Evidence of magnetic field switch-off in collisionless magnetic
  reconnection.
\newblock {\em The Astrophysical Journal Letters}, 810(2):L19, 2015.

\bibitem{lapenta2017}
Giovanni Lapenta.
\newblock Exactly energy conserving semi-implicit particle in cell formulation.
\newblock {\em Journal of Computational Physics}, 334:349--366, 2017.

\bibitem{lapenta2017b}
Giovanni Lapenta, Diego Gonzalez-Herrero, and Elisabetta Boella.
\newblock Multiple-scale kinetic simulations with the energy conserving
  semi-implicit particle in cell method.
\newblock {\em Journal of Plasma Physics}, 83(2), 2017.

\bibitem{lapenta2014}
Giovanni Lapenta, Stefano Markidis, Andrey Divin, David Newman, and Martin
  Goldman.
\newblock Separatrices: the crux of reconnection.
\newblock {\em arXiv preprint arXiv:1406.6141}, 2014.

\bibitem{lapenta2015}
Giovanni Lapenta, Stefano Markidis, Martin~V Goldman, and David~L Newman.
\newblock Secondary reconnection sites in reconnection-generated flux ropes and
  reconnection fronts.
\newblock {\em Nature Physics}, 11(8):690, 2015.

\bibitem{markidis2010}
Stefano Markidis, Giovanni Lapenta, et~al.
\newblock Multi-scale simulations of plasma with ipic3d.
\newblock {\em Mathematics and Computers in Simulation}, 80(7):1509--1519,
  2010.

\bibitem{Mazouffre2016}
Stéphane Mazouffre.
\newblock Electric propulsion for satellites and spacecraft: established
  technologies and novel approaches.
\newblock {\em Plasma Sources Science and Technology}, 25(3):033002, 2016.

\bibitem{olshevsky2015b}
V.~Olshevsky, J.~Deca, A.~Divin, I.B. Peng, S.~Markidis, M.E. Innocenti,
  E.~Cazzola, and G.~Lapenta.
\newblock Magnetic null points in kinetic simulations of space plasmas.
\newblock {\em arXiv preprint arXiv:1512.02018}, 2015.

\bibitem{priest2007}
Eric Priest and Terry Forbes.
\newblock Magnetic reconnection.
\newblock {\em Magnetic Reconnection, by Eric Priest, Terry Forbes, Cambridge,
  UK: Cambridge University Press, 2007}, 1, 2007.

\bibitem{tronci2015}
Cesare Tronci and Enrico Camporeale.
\newblock Neutral vlasov kinetic theory of magnetized plasmas.
\newblock {\em Physics of Plasmas}, 22(2):020704, 2015.

\bibitem{turner2008}
Martin~JL Turner.
\newblock {\em Rocket and spacecraft propulsion: principles, practice and new
  developments}.
\newblock Springer Science \& Business Media, 2008.

\bibitem{wan2008c}
Weigang Wan and Giovanni Lapenta.
\newblock Electron self-reinforcing process of magnetic reconnection.
\newblock {\em Physical review letters}, 101(1):015001, 2008.

\bibitem{wan2008}
Weigang Wan and Giovanni Lapenta.
\newblock Micro-macro coupling in plasma self-organization processes during
  island coalescence.
\newblock {\em Physical review letters}, 100(3):035004, 2008.

\end{thebibliography}

\end{document}